\documentclass[a4paper]{JHEP3}
\usepackage[english]{babel}
\usepackage{amsmath,amssymb}
\usepackage[pdftex]{graphicx}
\usepackage{amsfonts}
\usepackage{bbm}
\usepackage{cite}
\newcommand{\bc}{\begin{center}}
\newcommand{\ec}{\end{center}}
\newcommand{\bt}{\begin{tabular}}
\newcommand{\et}{\end{tabular}}
\newcommand{\be}{\begin{equation}}
\newcommand{\ee}{\end{equation}}
\newcommand{\bea}{\begin{eqnarray}}
\newcommand{\eea}{\end{eqnarray}}
\newcommand{\bfig}{\begin{figure}}
\newcommand{\efig}{\end{figure}}
\newcommand{\ba}{\begin{array}}
\newcommand{\ea}{\end{array}}
\newcommand{\bi}{\begin{itemize}}
\newcommand{\ei}{\end{itemize}}
\newcommand{\ie}{{\it i.e. }}
\newcommand{\eg}{{\it e.g. }}
\newcommand{\bfr}{\begin{flushright}}
\newcommand{\efr}{\end{flushright}}
\newcommand{\bfl}{\begin{flushleft}}
\newcommand{\efl}{\end{flushleft}}

\def\calo{{\cal O}}

\def\e{{\rm e}}
\preprint{ULB-TH/10-29}
\title{Neutrino masses and the Cabibbo expansion}
\author{Fu-Sin Ling$^{\hbox{a}}$\\
$^{\hbox{a}}$Service de Physique Th\'eorique, Universit\'e Libre de Bruxelles, \\
Campus de la Plaine CP225, Bd du Triomphe, 1050 Brussels, Belgium\\
E-mail: \email{fling@ulb.ac.be}}
\abstract{We adopt a bottom-up approach to reconstruct the possible neutrino mass matrix patterns,
in the hypothesis that $\lambda = \sin \theta_C$, where $\theta_C$ is the Cabibbo angle, is a useful
expansion parameter in the neutrino sector as it is in the quark sector.
A numerical analysis of these Cabibbo structures shows that an inverted hierarchy spectrum is statistically favored by the neutrino oscillation data.
We then discuss how apparently conflicting mass and mixing data for quarks, charged leptons and neutrinos can fit in a unified perspective.
We show that large mixing angles can appear through the seesaw mechanism, even if the neutrino Dirac mass matrix has hierarchical eigenvalues and small mixing angles.
Finally, it is shown that the form of the heavy Majorana mass matrix may point towards theories with extra-dimensions.}


\begin{document}

%
\section{Introduction}
\label{sec:intro}

The phenomenon of neutrino oscillations has proven that neutrinos have tiny masses and large flavor mixings,
and this is the most solid experimental evidence of physics beyond the Standard Model.
When looking at the mass and mixing data for the elementary particles,
one is immediately faced with the apparent discrepancy of patterns that distinguish 
neutrinos from the other fermions of the Standard Model (SM).
Both the quarks and charged leptons sectors are characterized by steep mass hierarchies, from the
light electron with a mass of $m_e \simeq 511$~keV to the heavy top quark with a mass of $m_t \simeq 172$~GeV.
Also, the Cabibbo-Kobayashi-Maskawa (CKM) matrix which describes mixings in the quarks sector only 
contains small mixing angles, a feature that is naturally consistent with the presence of very
hierarchical masses. 

On the other hand, the flavor structure in the neutrino sector looks quite different.
Unlike the CKM matrix in the quark sector, the Pontecorvo-Maki-Nakagawa-Sakata (PMNS or MNS) matrix which relates 
neutrino flavor eigenstates to their mass eigenstates harbors two large mixing angles
seen in solar and atmospheric neutrino oscillation experiments,
together with a rather mild mass hierarchy, if any.
The yet unknown absolute values of the neutrino masses add themselves a further puzzle to the 
mass hierarchy, as they do not exceed a few eV, and are therefore much lighter than
any other fermion of the Standard Model.
To understand these puzzles and the disparity between quarks, charged leptons and neutrinos
in a unified framework certainly represents one of the biggest challenge in particle physics, 
as part of a more general flavor problem.

In this paper, we adopt a bottom-up approach to find which mass patterns may lead naturally to the 
observed data. This path will serve as a guide towards a possible well-motivated extension
of the Standard Model, for which experimental consequences and predictions can be derived.
So in Sec.~\ref{sec:data}, we reconstruct possible structures for the light neutrino effective Majorana mass matrix.
Then, in Sec.~\ref{sec:num}, a numerical analysis enables to test these structures against the experimental data.
In Sec.~\ref{sec:theory}, we address the question of a possible origin for these structures, in particular within
the seesaw mechanism framework. We also confront and find a solution to the apparent conflict between quarks, charged leptons
and neutrinos mass and mixing data, before concluding in Sec.~\ref{sec:end}.

\section{The flavor puzzle}
\label{sec:data}   

Let us start by writing down the mixing matrix in the leptonic sector, in the (minimal) three neutrino scheme.
The MNS matrix which relates neutrino flavor eigenstates to mass eigenstates, 
$\nu_f = U_{MNS} \; \nu_m$, can be decomposed as $U_{MNS} = U \cdot K$,
where $U$ is conventionally parametrized in the following form~\cite{Amsler:2008zzb}
\be
U(\theta_{12},\theta_{13},\theta_{23},\delta_\text{CP}) = \left(
\ba{ccc}
    c_{12} c_{13}
    & s_{12} c_{13}
    & s_{13} e^{-i\delta_\text{CP}}
    \\
    - s_{12} c_{23} - c_{12} s_{13} s_{23} e^{i\delta_\text{CP}}
    & \hphantom{+} c_{12} c_{23} - s_{12} s_{13} s_{23} e^{i\delta_\text{CP}}
    & c_{13} s_{23} \hspace*{5.5mm}
    \\
    \hphantom{+} s_{12} s_{23} - c_{12} s_{13} c_{23} e^{i\delta_\text{CP}}
    & - c_{12} s_{23} - s_{12} s_{13} c_{23} e^{i\delta_\text{CP}}
    & c_{13} c_{23} \hspace*{5.5mm}
\ea
\right) \quad ,
\label{eq:umns}
\ee
where $s_{ij} = \sin \theta_{ij}$, $c_{ij} = \cos \theta_{ij}$, $\delta_\text{CP}$ is a CP violating phase,
and $K={\rm diag}(1,\e^{i \phi_1},\e^{i \phi_2})$ is a diagonal matrix containing 
two additional Majorana phases which do not affect flavor oscillations.
The latest update on a three neutrino global fit gives
$\tan^2 \theta_{12} = 0.47_{-0.10}^{+0.14}$, $\tan^2 \theta_{23} = 0.9^{+1.0}_{-0.4}$,
$\sin^2 \theta_{13} \leq 0.05$ at $3\sigma$ C.L with a possible hint at $1\sigma$ that $\theta_{13}>0$.
The $CP$ phase is left unconstrained, and the mass squared splittings $\Delta m_{ij}^2 = m^2_{\nu_i}-m^2_{\nu_j}$
are determined as $\Delta m^2_{21} = 7.6 \pm 0.7 \times 10^{-5}~{\rm eV^2}$,
$\Delta m^2_{31} = 2.46 \pm 0.37 \times 10^{-3}~{\rm eV^2}$ (normal scheme) or
$\Delta m^2_{31} = -2.36 \pm 0.37 \times 10^{-3}~{\rm eV^2}$ (inverted
scheme)~\cite{GonzalezGarcia:2010er}.
The MNS matrix harbors two large mixing angles $\theta_{12}$ and $\theta_{23}$ that are seen respectively 
in solar and atmospheric neutrino oscillation experiments. The third angle $\theta_{13}$, usually called the 
reactor angle as the best limit comes from the nuclear reactor experiment CHOOZ~\cite{Apollonio:2002gd}. 
The mass hierarchy in the neutrino sector is rather mild, as $|\Delta m^2_{21}/\Delta m^2_{31}| \simeq3 \%$.
It is even possible that there is no mass hierarchy at all, if the three neutrinos are degenerate in mass,
with hyperfine splittings. While the absolute mass scale for the neutrinos is still unknown,
bounds from cosmology~\cite{Komatsu:2010fb} and from beta-decay experiments~\cite{Kraus:2004zw}
show that they are in any case much lighter than any other fermion in the Standard Model. 
The latest bound, based on numerical predictions of the effect of neutrino masses on the clustering 
of small-scale structures gives $\sum m_{\nu_i} < 0.28$~eV at $95\%$~C.L.~\cite{Thomas:2009ae}.

These new observations in the neutrino phenomenology are in stark contrast with known masses and mixings for
quarks and charged leptons. Standard Model fermions come in three families of increasing mass (except for neutrinos ?),
with both the number of generations and the mass hierarchy within one family or between different families
being a complete mystery. Moreover, the mixing matrix in the quark sector only exhibits small angles.
With the parameterization of Eq.~(\ref{eq:umns}) for the CKM matrix $V$ (or $V_{CKM}$), the latest determination gives  
$\theta_{12} = 13.04 \pm 0.05^\circ$, $\theta_{13} = 0.201 \pm 0.011^\circ$, $\theta_{23} = 2.38 \pm 0.06^\circ$
and $\delta_{CP} = 1.20 \pm 0.08$~\cite{Amsler:2008zzb}.

Clearly, neutrinos add a further challenge to the flavor puzzle. 
Several important questions also need to be answered in the future: the nature of the neutrino mass (Dirac or Majorana), 
their absolute mass scale, the sign of the hierarchy, the size of the reactor angle which also controls 
the size of a possible $CP$ violation in the leptonic sector.
Altogether, these clues represent paths (or maybe highways) to the correct theory beyond the Standard Model,
therefore it is worthwhile to reflect on what we have already learned from them.

\subsection{Cabibbo expansion}

The smallness of the mixing angles in $V_{CKM}$ leads to a useful approximation known as the Wolfenstein 
parameterization
\be
V_{CKM} = \left(
\ba{ccc}
    1 - \lambda^2/2 & \lambda & A \lambda^3 (\rho - i \eta)\\
    - \lambda & 1 - \lambda^2/2 & A \lambda^2\\
    A \lambda^3 (1 - \rho - i \eta) & - A \lambda^2 & 1
\ea
\right) + \calo(\lambda^4) \quad ,
\label{eq:wolf}
\ee
where $\lambda = \sin \theta_C$ serves as an expansion parameter, and $\theta_C \simeq 13^\circ$ is the Cabibbo angle. 
All the Wolfenstein parameters have order one values, the best determination being
$\lambda = 0.2257^{+0.0009}_{-0.0010}$, $A = 0.814^{+0.021}_{-0.022}$, 
$\rho = 0.135^{+0.031}_{-0.016}$, and $\eta = 0.349^{+0.015}_{-0.017}$.

The fact that $|V_{td}| \sim |V_{ts}|\cdot|V_{us}|$ and several empirical relations like
$|V_{us}| \simeq \sqrt{m_d/m_s}$ (known as the Gatto-Sartori-Tonin/Cabibbo-Maiani 
relation~\cite{Gatto:1968ss, Cabibbo:1968vn}) or $|V_{ub}/V_{cb}| \simeq \sqrt{m_u/m_c}$
advocate for a direct connection between quark masses and their mixings.
As an example of a simple realization, we can mention the Froggatt-Nielsen mechanism~\cite{Froggatt:1978nt}.
The idea is to reproduce the observed fermion mass hierarchies by introducing Abelian family symmetries 
which are broken at high scale by the vacuum expectation value of some flavon field, and by assigning 
suitable charges to the Standard Model fields. 
For example, in Ref.~\cite{Ling:2002nj}, the quark matrices have the following Cabibbo structure
\be
M_u \sim \left(
\ba{ccc}
    \lambda^6 & \lambda^5 & \lambda^3\\
    \lambda^5 & \lambda^4 & \lambda^2\\
    \lambda^3 & \lambda^2 & 1
\ea
\right) \quad , \quad
M_d \sim \left(
\ba{ccc}
    \lambda^4 & \lambda^3 & \lambda^3\\
    \lambda^3 & \lambda^2 & \lambda^2\\
    \lambda & 1 & 1
\ea
\right) \quad .
\label{eq:mumd}
\ee
Assuming $\calo(1)$ numerical prefactors, the diagonalization of these matrices, $M_u = U_u \tilde{M_u} V_u^\dagger$ and 
$M_d = U_d \tilde{M_d} V_d^\dagger$, leads automatically to the correct structure for the 
CKM matrix $V_{CKM} = U_u^\dagger U_d$, explicit in the parametrization Eq.~(\ref{eq:wolf}).
The bottom line is that for quarks, hierarchical masses and small mixing angles fit in a coherent 
picture where the structure of the mass matrices is set by the leading power of the Cabibbo
parameter for each element, assuming prefactor coefficients of order one.

Can this scheme be extended to leptons, and in particular to the neutrino sector 
where large mixing angles are found ?
First of all, we don't know whether neutrinos masses are of Dirac or Majorana type.
However, the seesaw mechanism~\cite{Ramond:1979py,Yanagida:1979as} gives a compelling explanation for the smallness of the 
neutrino masses. Also, as we will argue, it provides subtle ways to generate large mixing angles.
Therefore, in this paper, it will be assumed that light neutrinos have Majorana masses.
Secondly, in full generality, $U_{MNS} = U_l^\dagger U_\nu$ is a product
of two matrices $U_l$ and $U_\nu$ coming from the diagonalization of 
$M_l = U_l \tilde{M_l} V_l^\dagger$ and $M_\nu = U_\nu \tilde{M_\nu} U_\nu^t$.
Therefore the two large mixing angles found in $U_{MNS}$ can possibly be shared between the 
charged lepton and the neutrino sectors. For example, in the Froggatt-Nielsen framework,
the structure of $M_d$ in Eq.~(\ref{eq:mumd}) can be suggestive of one large mixing angle in $M_l$, 
as $M_l \sim M_d^\dagger$ are related in a Grand Unification perspective. 

In the following, we will therefore consider two situations.
In the first situation, the two large angles in $U_{MNS}$ originate from the neutrino mass matrix alone,
while the charged lepton sector contains only small mixing angles, and we take $U_l \simeq V_{CKM}$.
In the second situation, we attribute the large atmospheric and the large solar angles
to the charged lepton and the neutrino sectors respectively.
In each case, the different possible Cabibbo structures of the mass matrix are reconstructed, in order to investigate
whether the observed oscillation data can be understood in terms of such structures.

\subsection{Neutrino mass matrix when $U_\nu$ contains two large mixing angles}
\label{sec:22}

The goal of this section is to reconstruct the different possible Cabibbo structures of the light neutrino 
effective Majorana mass matrix that are compatible with the neutrino oscillation data.
In this perspective, we will view the smallness of the ratio 
\be
\Delta m^2_{21}/\Delta m^2_{31} \sim \lambda^2
\label{eq:dm2r}
\ee
as a remnant of a hierarchical pattern in the neutrino sector. 
In this section, we consider the case where both large mixing angles come from $U_\nu$. 
We denote the mixing angles in $U_\nu$ by $\phi_{12}$, $\phi_{23}$, and $\phi_{13}$.
Taking into account possible small rotations coming the charged lepton sector, with $U_l \simeq V_{CKM}$, 
we consider that both $\phi_{12}$ and $\phi_{23}$ are large, and can even be maximal, while $\phi_{13}$ is smaller
than around $10^\circ$. \\

\underline{Normal hierarchy}\\ 

If the hierarchy of the light neutrino masses is normal, we have 
$(m_{\nu_1},m_{\nu_2},m_{\nu_3}) \sim m_{\nu_3} \cdot (\lambda^\alpha,\lambda,1)$ with $\alpha \geq 1$, 
so that the relation Eq.~(\ref{eq:dm2r}) is \emph{a priori} satisfied.
This leads to a reconstructed matrix $M_\nu$ with a dominant $2-3$ block
\be
M_\nu \sim \left(
\ba{ccc}
    \lambda & \lambda & \lambda\\
    \lambda & 1 & 1 \\
    \lambda & 1 & 1
\ea
\right) \quad .
\label{eq:nh}
\ee
The drawback is that this structure does not automatically lead to the appearance of large mixing angles. 
Even worse, it does not automatically lead to hierarchical neutrino masses 
$m_{\nu_1} \leq m_{\nu_2} \ll m_{\nu_3}$ nor to the relation Eq.~(\ref{eq:dm2r}) as we supposed in the beginning.
To ensure this, we need to impose a supplementary condition on the subdeterminant $D_1$ of the $2-3$ block,
namely $D_1 \sim \lambda^\beta$ with $\beta \geq 1$~\cite{Ling:2002nj}. 
While the condition on $D_1$ might appear as a fine-tuning requirement for the numerical prefactors, 
it can actually be achieved quite naturally in the context of the seesaw mechanism,
in the case of a single right-handed neutrino dominance for example~\cite{King:2002nf}.

Unfortunately, the matrix structure Eq.~(\ref{eq:nh}) generically predicts small mixing angles
$\theta_{12}$ and $\theta_{13}$. To get a large angle $\theta_{12}$, a strong fine-tuning is needed, 
and the seesaw mechanism does not bring any improvement in this respect, 
as we will show in our numerical study. \\

\underline{Inverted hierarchy} \\

In the inverted hierarchy case, we take $|m_{\nu_2}| = |m_{\nu_1}|(1+a \lambda^2)$ where $a>0$ is a 
$\calo(1)$ numerical coefficient, and $m_{\nu_3} \sim \lambda^\alpha$ with $\alpha \geq 1$,
so that the relation Eq.~(\ref{eq:dm2r}) is recovered.
The reconstructed matrix $M_\nu$ does not exhibit any interesting Cabibbo structure,
except for one particular case, when $m_{\nu_2} \simeq -m_{\nu_1}$ and the angle $\phi_{12}$ is maximal,
\be
M_\nu \sim \left(
\ba{ccc}
    \lambda^2 & 1 & 1 \\
    1 & \lambda^2 & \lambda^2 \\
    1 & \lambda^2 & \lambda^2
\ea
\right) \quad .
\label{eq:ih}
\ee
The leading entries in this matrix are off-diagonal. 
The other elements are at most of order $\lambda^2$ when $\lambda \simeq 0.1 \dots 0.2$, which ensures that 
the ratio of $\Delta m^2$ obeys Eq.~(\ref{eq:dm2r}) and also that the lightest neutrino mass is of order $\lambda$.
The Cabibbo structure Eq.~(\ref{eq:ih}) leads to a maximal angle 
$\phi_{12} = \pi/4 + n \pi/2 +\calo(\lambda^2)$ ($n$ integer), a large but not necessarily maximal angle $\phi_{23}$,
and a small angle $\phi_{13} \sim \lambda$.

As we will show in the numerical study Sec.~\ref{sec:num}, the Cabibbo structure Eq.~(\ref{eq:ih}) can easily account for the 
observed data in neutrino oscillations experiments, once small mixing angles from the charged lepton sector are taken into
account. No fine-tuning in the numerical coefficients is needed. Also large mixing angles appear without the introduction
of any additional discrete symmetry. \\

\underline{Degenerate case} \\

In the quasi degenerate case, we take $|m_{\nu_2}| = |m_{\nu_1}|(1 + a \lambda^3)$ 
and $|m_{\nu_3}| = |m_{\nu_1}|(1 + b \lambda)$, where $a>0$ and $b$ are $\calo(1)$ numerical coefficients,
in order to satisfy the relation Eq.~(\ref{eq:dm2r}).
The reconstructed matrix $M_\nu$ does not exhibit any interesting Cabibbo structure,
except for one particular case, when $m_{\nu_2} \simeq m_{\nu_1}$, $m_{\nu_3} \simeq -m_{\nu_1}$ 
and $U_\nu$ is bimaximal ($\phi_{12}$ and $\phi_{23}$ are maximal),
\be
M_\nu \sim \left(
\ba{ccc}
    1 & \lambda^3 & \lambda^3 \\
    \lambda^3 & \lambda & 1 \\
    \lambda^3 & 1 & \lambda
\ea
\right) \quad .
\label{eq:dh}
\ee
However, a mass matrix with the structure Eq.~(\ref{eq:dh}) but with random $\calo(1)$ coefficients
does not lead in general to two large and one small angles, and does not lead to a quasi-degenerate spectrum.
Moreover, the ratio $\Delta m^2_{21}/\Delta m^2_{31}$ is not necessarily small.
As the numerical study Sec.~\ref{sec:num} shows, it is almost impossible to accommodate 
the experimental data and obtain quasi degenerate neutrinos from a Cabibbo structure like Eq.~(\ref{eq:dh}) 
without imposing relations among the numerical prefactors.

\subsection{Neutrino mass matrix when $U_\nu$ contains one large mixing angle}
\label{sec:23}

When the large atmospheric angle is attributed to the charged lepton sector, there are two possible decompositions 
of $U_{MNS} = U_l^\dagger U_\nu$ to zeroth order in the Cabibbo expansion parameter $\lambda$, namely
\be
U_l = \left(
\ba{ccc}
    1 & 0 & 0 \\
    0 & \cos \theta_\oplus & \sin \theta_\oplus \\
    0 & -\sin \theta_\oplus & \cos \theta_\oplus
\ea
\right) + \calo(\lambda) \quad ,
\label{eq:ul}
\ee
and
\be
U_\nu = \left(
\ba{ccc}
    \cos \theta_\odot & \sin \theta_\odot & 0 \\
    -\sin \theta_\odot & \cos \theta_\odot & 0 \\
	0 & 0 & 1
\ea
\right) + \calo(\lambda) \quad 
\label{eq:unu1}
\ee
or
\be
U_\nu = \left(
\ba{ccc}
    \cos \theta_\odot & \sin \theta_\odot & 0 \\
	0 & 0 & 1 \\
    -\sin \theta_\odot & \cos \theta_\odot & 0 \\
\ea
\right) + \calo(\lambda) \quad ,
\label{eq:unu2}
\ee
where $\theta_\oplus$ and $\theta_\odot$ are the atmospheric and solar angles respectively.
Again, we reconstruct the neutrino mass matrix structure in the normal, inverted and degenerate hierarchy cases. \\

\underline{Normal hierarchy}\\ 

When the neutrino masses are hierarchical 
$(m_{\nu_1},m_{\nu_2},m_{\nu_3}) \sim m_{\nu_3} \cdot (\lambda^\alpha,\lambda,1)$ with $\alpha \geq 1$, 
and $U_\nu$ contains only one large angle Eq.~(\ref{eq:unu1}) or Eq.~(\ref{eq:unu2}), we get
\be
M_\nu \sim \left(
\ba{ccc}
    \lambda & \lambda & \lambda\\
    \lambda & \lambda & \lambda \\
    \lambda & \lambda & 1
\ea
\right) \quad {\rm or} \quad
M_\nu \sim \left(
\ba{ccc}
    \lambda & \lambda & \lambda\\
    \lambda & 1 & \lambda \\
    \lambda & \lambda & \lambda
\ea
\right) \quad .
\label{eq:nh2}
\ee
The advantage of these structures compared to Eq.~(\ref{eq:nh}) is that no condition on a subdeterminant is required.
However, the large angle is not predicted but merely accidental.
A particular case arises if $\theta_\odot$ is maximal and $m_{\nu_1} \simeq - m_{\nu_2}$, we get
\be
M_\nu \sim \left(
\ba{ccc}
    \lambda^2 & \lambda & \lambda\\
    \lambda & \lambda^2 & \lambda \\
    \lambda & \lambda & 1
\ea
\right) \quad {\rm or} \quad
M_\nu \sim \left(
\ba{ccc}
    \lambda^2 & \lambda & \lambda\\
    \lambda & 1 & \lambda \\
    \lambda & \lambda & \lambda^2
\ea
\right) \quad .
\label{eq:nh3}
\ee
Interestingly, these two structures do lead automatically to a large solar angle,
with a hierarchical mass pattern.\\

\underline{Inverted hierarchy} \\

The inverted hierarchy of the neutrino masses can only be guaranteed by two structures,
\be
M_\nu \sim \left(
\ba{ccc}
    \lambda^2 & 1 & \lambda\\
    1 & \lambda^2 & \lambda \\
    \lambda & \lambda & \lambda
\ea
\right) \quad {\rm or} \quad
M_\nu \sim \left(
\ba{ccc}
    \lambda^2 & \lambda & 1\\
    \lambda & \lambda & \lambda \\
    1 & \lambda & \lambda^2
\ea
\right) \quad ,
\label{eq:ih2}
\ee
which lead to a solar angle close to its maximal value, 
$|m_{\nu_2}| = |m_{\nu_1}|(1+a \lambda^2)$ where $a>0 \sim \calo(1)$, and $m_{\nu_3} \sim \lambda$.
Notice that the powers $\lambda^2$ in position $(1,1)$ and $(2,2)$ (or $(3,3)$) are needed to guarantee 
the relation Eq.~(\ref{eq:dm2r}).\\

\underline{Degenerate case} \\

As in the previous section, no Cabibbo structure can lead to a quasi degenerate spectrum.
Let us mention that the following structures 
\be
M_\nu \sim \left(
\ba{ccc}
    \lambda^2 & 1 & \lambda\\
    1 & \lambda^2 & \lambda \\
    \lambda & \lambda & 1
\ea
\right) \quad {\rm or} \quad
M_\nu \sim \left(
\ba{ccc}
    \lambda^2 & \lambda & 1\\
    \lambda & 1 & \lambda \\
    1 & \lambda & \lambda^2
\ea
\right) \quad ,
\label{eq:dh2}
\ee
give an angle close to maximal in the $1-2$ or $1-3$ block and 
$m_{\nu_1} \simeq -m_{\nu_2} \sim m_{\nu_3}$ with a hierarchy of $\Delta m^2$
as required by Eq.~(\ref{eq:dm2r}).

\section{Numerical study} 
\label{sec:num}

In this section, the Cabibbo structures of the reconstructed neutrino mass matrix derived 
in Sec.~\ref{sec:data} are analyzed and compared numerically.
The goal is to see quantitatively whether these structures are likely to reproduce the observed 
data on a statistical basis. 
This numerical study can serve as a benchmark in the quest for a theory of neutrino masses.

So, we consider the structures of Eqs.~(\ref{eq:nh}-\ref{eq:dh}) and Eqs.~(\ref{eq:nh2}-\ref{eq:dh2}) for the neutrino 
mass matrix, with random $\calo(1)$ coefficients for each element.
As the neutrino effective Majorana mass matrix is symmetric, there are only six independent prefactors,
and we limit to real coefficients.
Apart from a random sign, every $\calo(1)$ coefficient $c$ is chosen as a power of $10$, 
so $c = \pm 10^X$, where $X$ is a random variable following a Gaussian distribution $N(0,\sigma)$,
with $\sigma = 0.3$. Therefore, at $99\%$ C.L., $0.169 \leq |c| \leq 5.93$.
We set the value of the expansion parameter to $\lambda = 0.1$.
For a given structure, a sample with size $N=10^5$ of randomly generated mass matrices is considered.

The diagonalization of the neutrino mass matrix yields $U_\nu$. 
The neutrino mass eigenstates are labeled $\nu_1$, $\nu_2$, $\nu_3$, in such a way that $m_{\nu_2}^2-m_{\nu_1}^2$
corresponds to the smallest positive mass squared difference.
To obtain the MNS matrix, we take into account a possible rotation coming from the charged lepton sector.
Following Sec.~\ref{sec:data}, we consider successively the two possibilities of zero or one large angle in $U_l$.
In the first case, both large angles found in $U_{MNS}$ stem from the neutrino sector, and 
we suppose $U_l \simeq V_{CKM}$ (but we drop the $CP$ phase) in a quark-lepton unified perspective.
So, in terms of the parameterization Eq.~(\ref{eq:umns}), we take $U_l = U(13.0^\circ,0.2^\circ,2.4^\circ,0)$. 
In the second case, we suppose that $U_l$ contains a maximal angle in the $2-3$ block, so 
we take $U_l = U(13.0^\circ,0.2^\circ,45.0^\circ,0)$. Such maximal angle could arise from
a charged lepton mass matrix with a Cabibbo structure
\be
M_l \sim \left(
\ba{ccc}
    \lambda^4 & \cdot & \cdot \\
    \cdot & \lambda^2 & 1 \\
    \cdot & \cdot & 1
\ea
\right) \quad ,
\ee
where the elements replaced with dots are suppressed enough in order for
the two other angles to remain small, and to keep hierarchical masses $\sim (\lambda^4,\lambda^2,1)$.
In this paper however, we will focus on the neutrino sector, and therefore leave aside the question of 
constructing structures in the charged lepton sector with a large angle. 

\subsection{$U_l$ contains zero large angle}
\label{sec:num1}

\hspace{0.5cm}
\underline{Normal hierarchy}\\

\begin{figure*}[t]
\begin{center}
\begin{tabular}{ccc}
\includegraphics[width=0.45 \textwidth, clip]{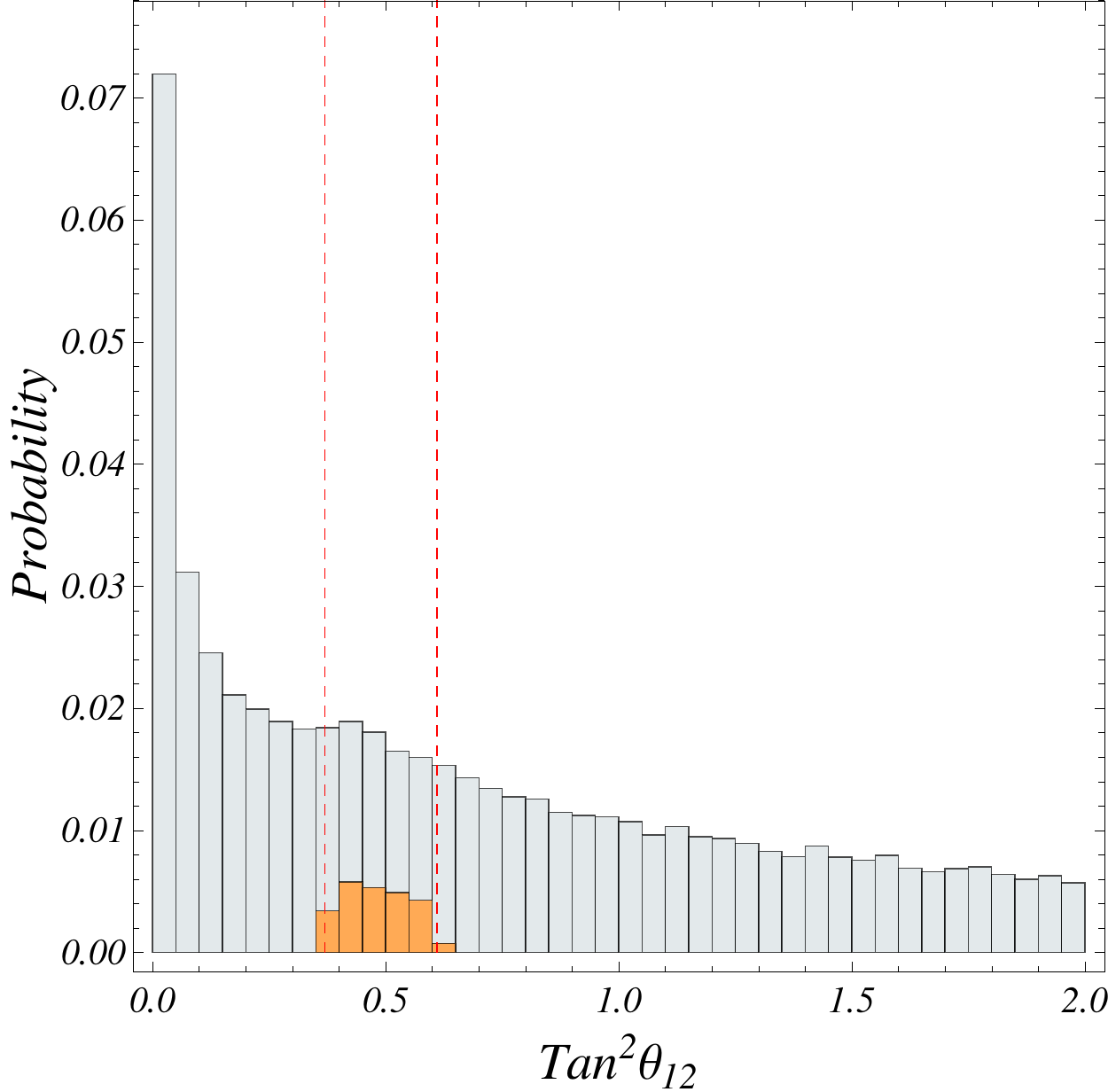}& ~ &
\includegraphics[width=0.45 \textwidth, clip]{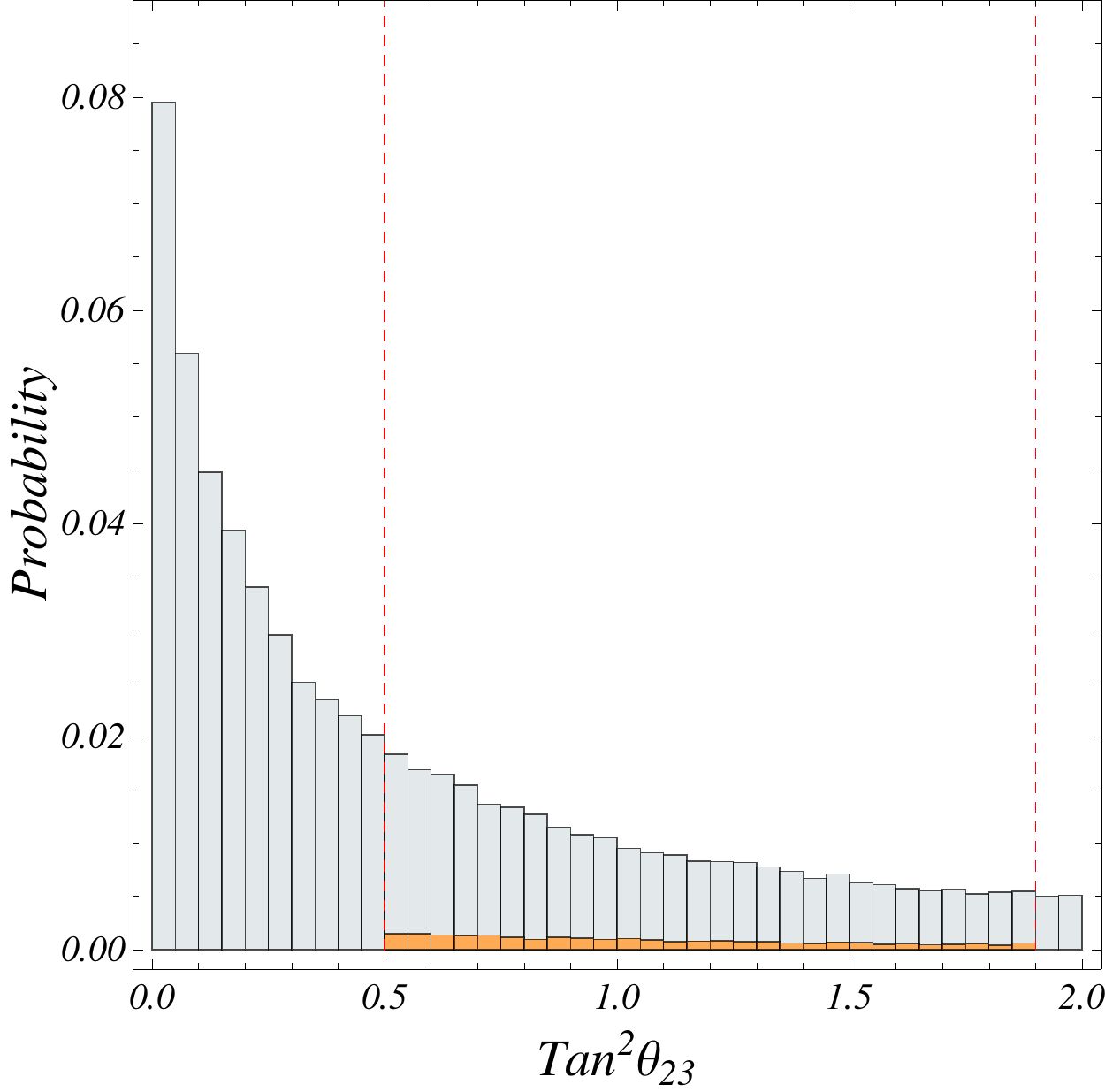}\\
\includegraphics[width=0.45 \textwidth, clip]{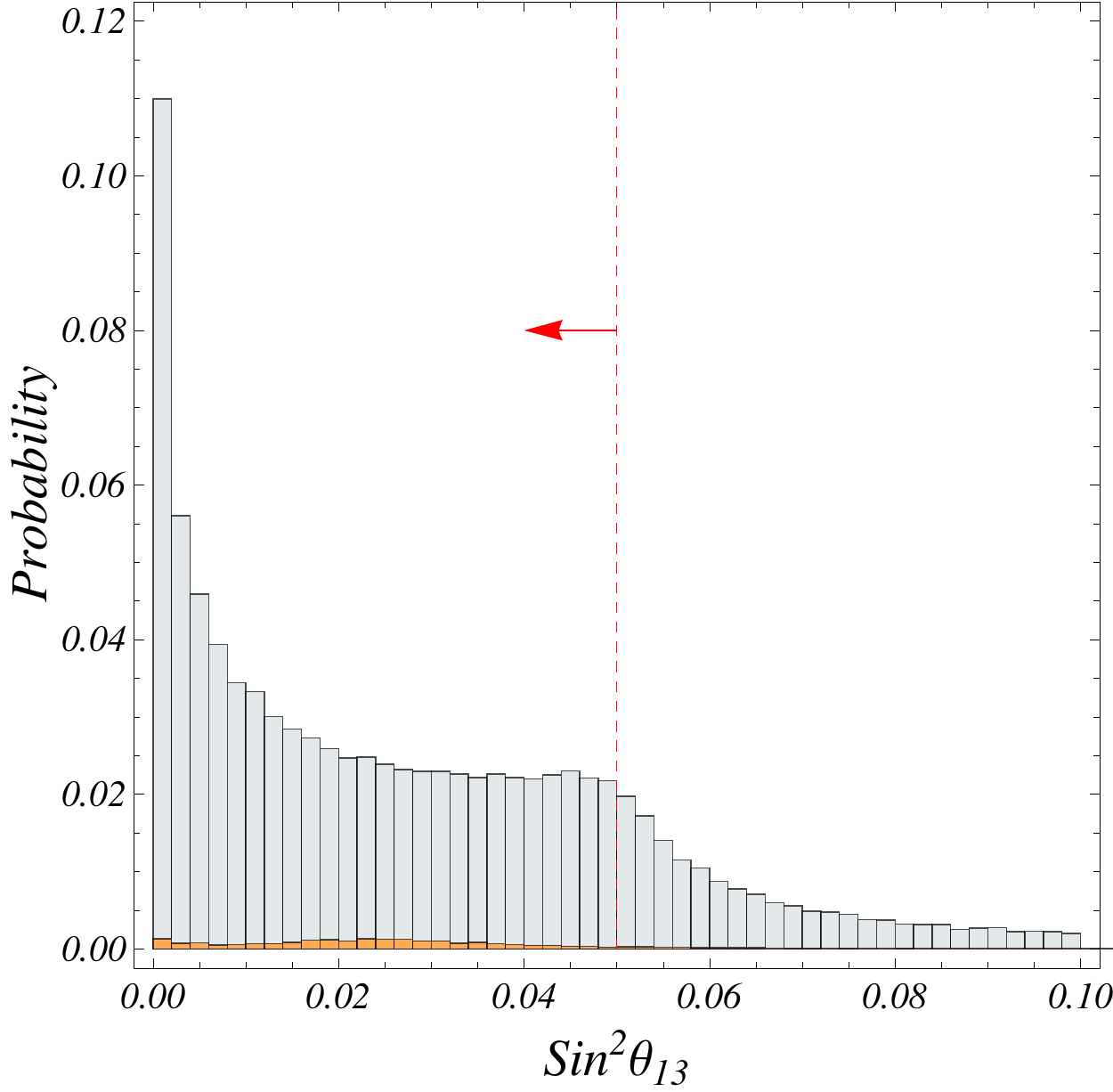}& ~ &
\includegraphics[width=0.45 \textwidth, clip]{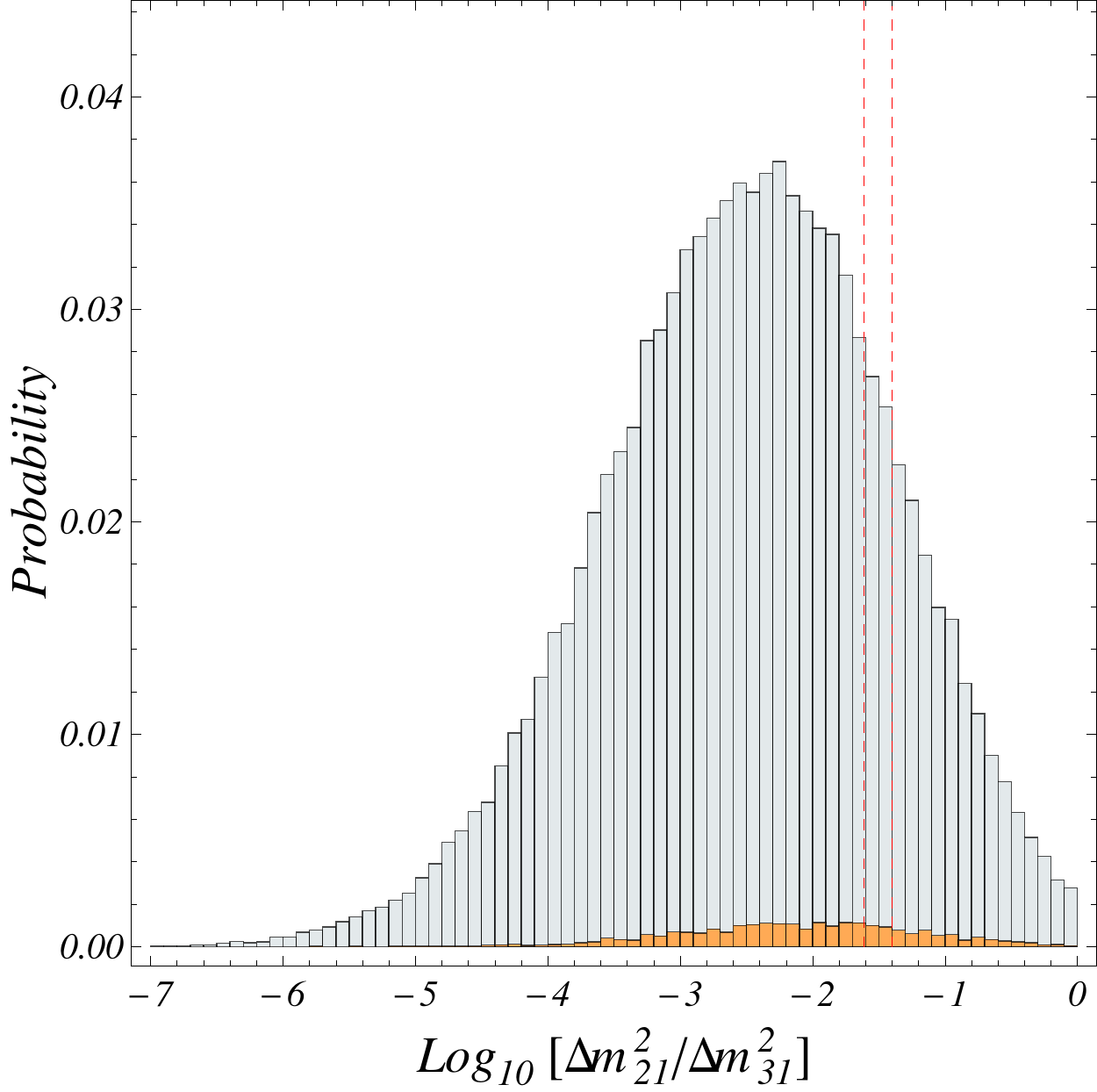}\\
\end{tabular}
\caption{\small \it 
{\rm \underline{Normal hierarchy}} :  
Distributions of $\tan^2 \theta_{12}$, $\tan^2 \theta_{23}$, $\sin^2 \theta_{13}$ and
$\Delta m^2_{21}/\Delta m^2_{31}$ when the neutrino mass matrix has the 
Cabibbo structure given by Eq.~(\ref{eq:nh}) with a subdeterminant in the $2-3$ block $D_1 \sim \lambda$,
and with $U_l = U(13.0^\circ,0.2^\circ,2.4^\circ,0)$ ($U_l \simeq V_{CKM}$). 
The size of the sample is $N=10^5$ and the expansion parameter is $\lambda=0.1$. 
The $\calo(1)$ numerical prefactors have the form $\pm 10^X$,
where $X$ is a random variable following a normal distribution $N(0,0.3)$.
In orange color : the distributions of the subsample ($2.5\%$ of total sample) for which 
$0.37 < \tan^2 \theta_{12} <0.61$ and $0.5 < \tan^2 \theta_{23} < 1.9$.
Dashed lines (in red) indicate the experimental range at $3\sigma$ C.L.
Only the central parts of the distributions are displayed, the tails have been cut.}
\label{fig:nh}
\end{center}
\end{figure*}
In the normal hierarchy case, with the structure Eq.~(\ref{eq:nh}), 
as we saw in Sec.~\ref{sec:22}, it is necessary to impose a condition
on the subdeterminant $D_1$ of the $2-3$ block of the matrix Eq.~(\ref{eq:nh}) to have hierarchical
neutrinos masses. Therefore, we parameterize this $2-3$ block as 
$(M_\nu)_{22} = a^2$, $(M_\nu)_{33} = b^2$, and $(M_\nu)_{23} = a b + c \lambda$, where $a,b,c$ are
$\calo(1)$ random coefficients.

The results are shown on Fig.~\ref{fig:nh}. 
Both angles $\theta_{12}$ and $\theta_{23}$ have a distribution with a wide extension,
and a high probability of being small. 
The probability that both angles $\theta_{12}$ and $\theta_{23}$ are large
and within the experimental range at $3\sigma$ C.L. (\ie $0.37 < \tan^2 \theta_{12} < 0.61$
and $0.5 < \tan^2 \theta_{23} < 1.9$) is only $2.5\%$. 
Also, the angle $\theta_{13}$ can be larger than the current bound (with a probability of around $20\%$), 
even when the two other angles are large. Finally, the hierarchy of the $\Delta m^2$ is also only 
broadly determined by the Cabibbo structure of the neutrino mass matrix.
The influence of the numerical prefactors is important, with a ratio $\Delta m^2_{21}/\Delta m^2_{31}$
ranging from $10^{-6}$ to $1$. \\

\underline{Inverted hierarchy} \\

\begin{figure*}[t]
\begin{center}
\begin{tabular}{ccc}
\includegraphics[width=0.45 \textwidth, clip]{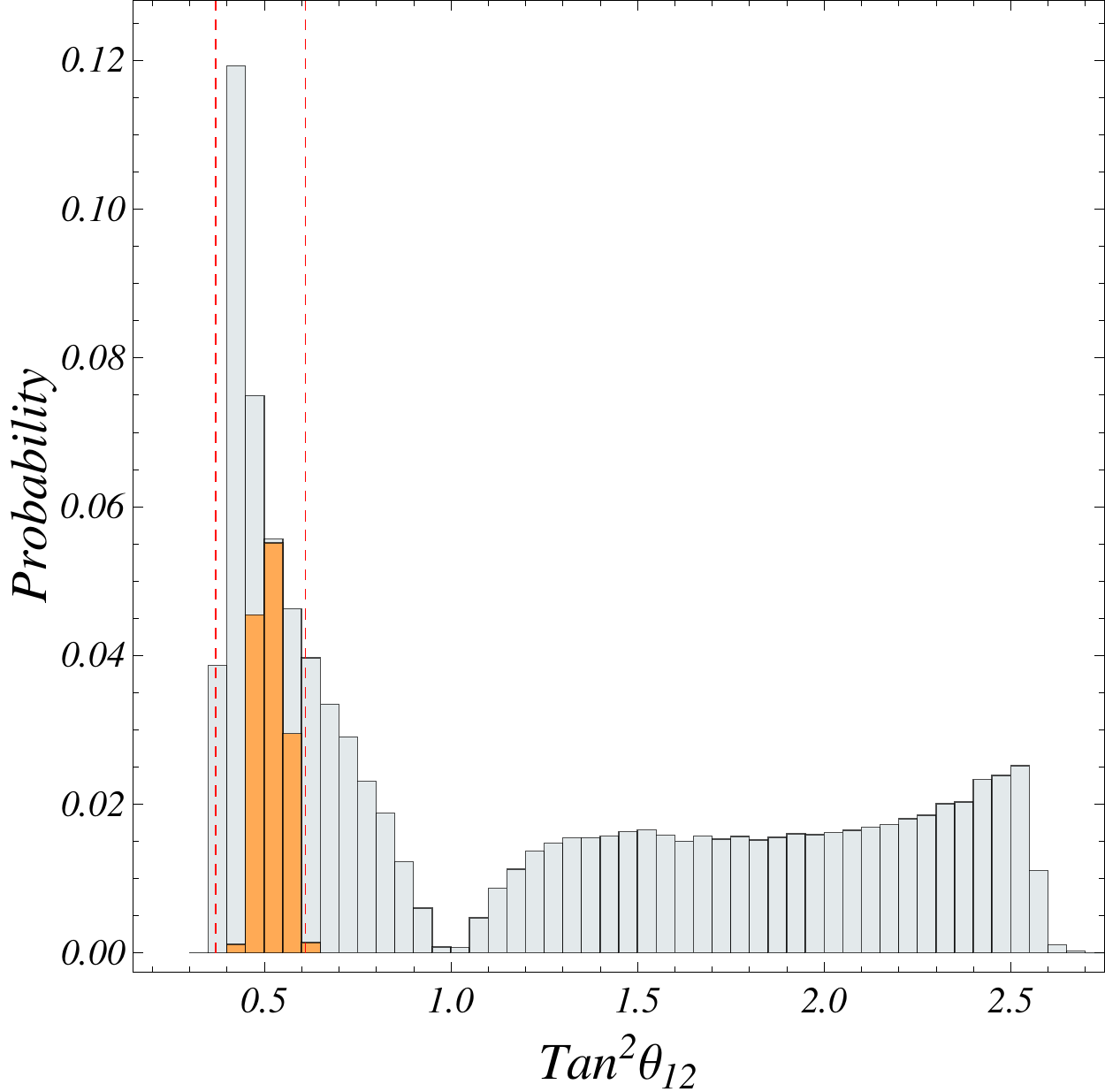}& ~ &
\includegraphics[width=0.45 \textwidth, clip]{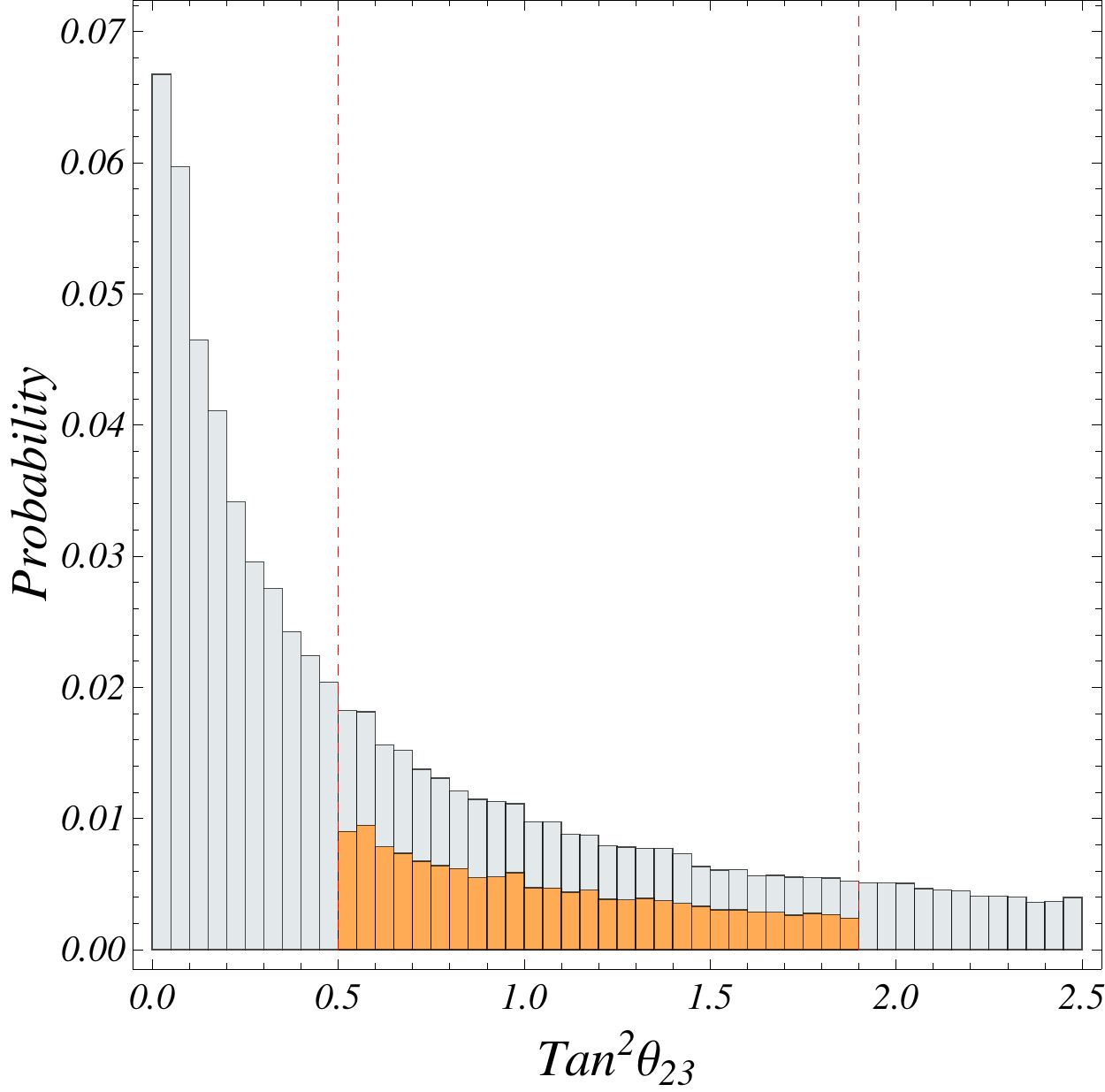}\\
\includegraphics[width=0.45 \textwidth, clip]{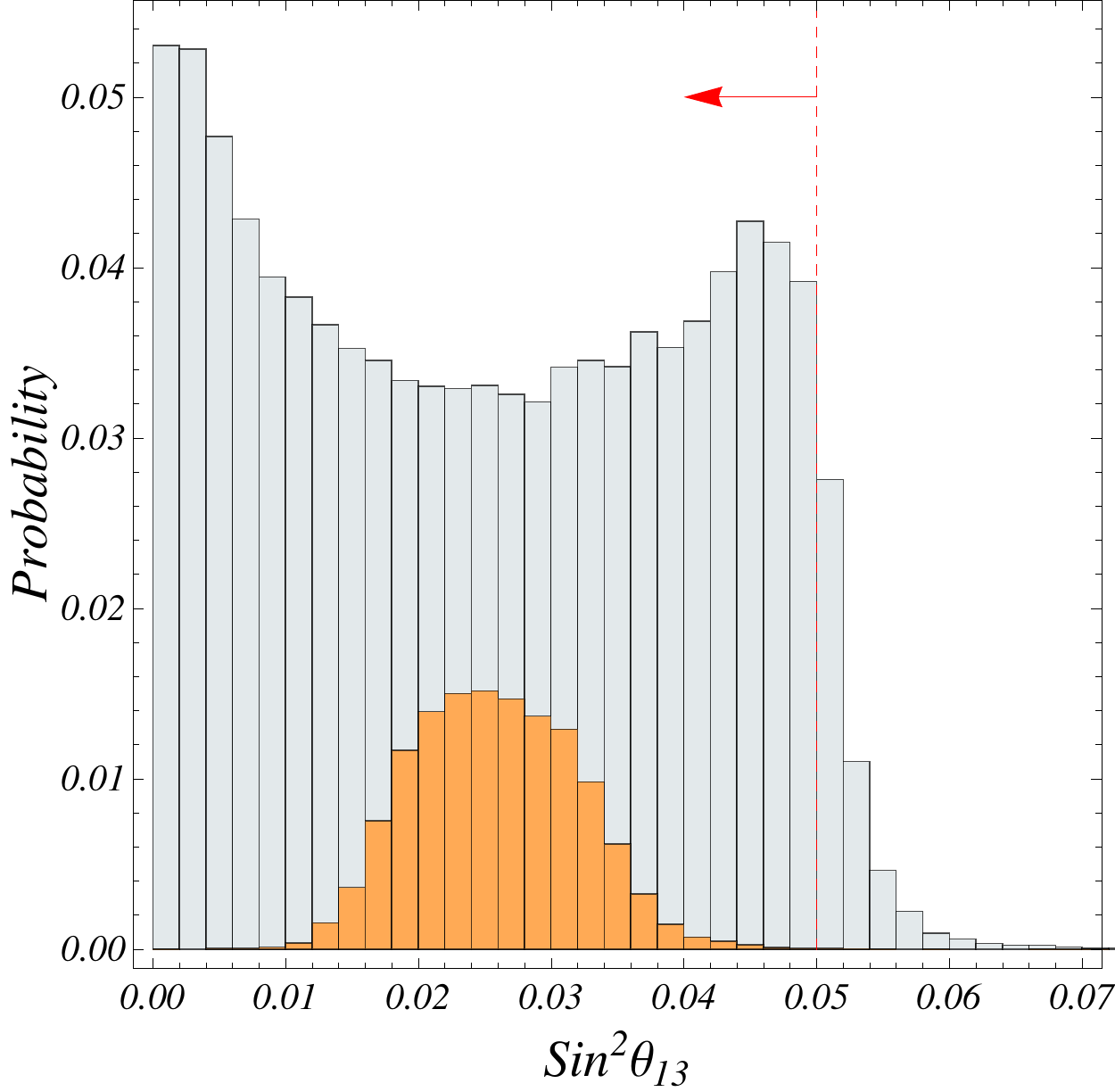}& ~ &
\includegraphics[width=0.45 \textwidth, clip]{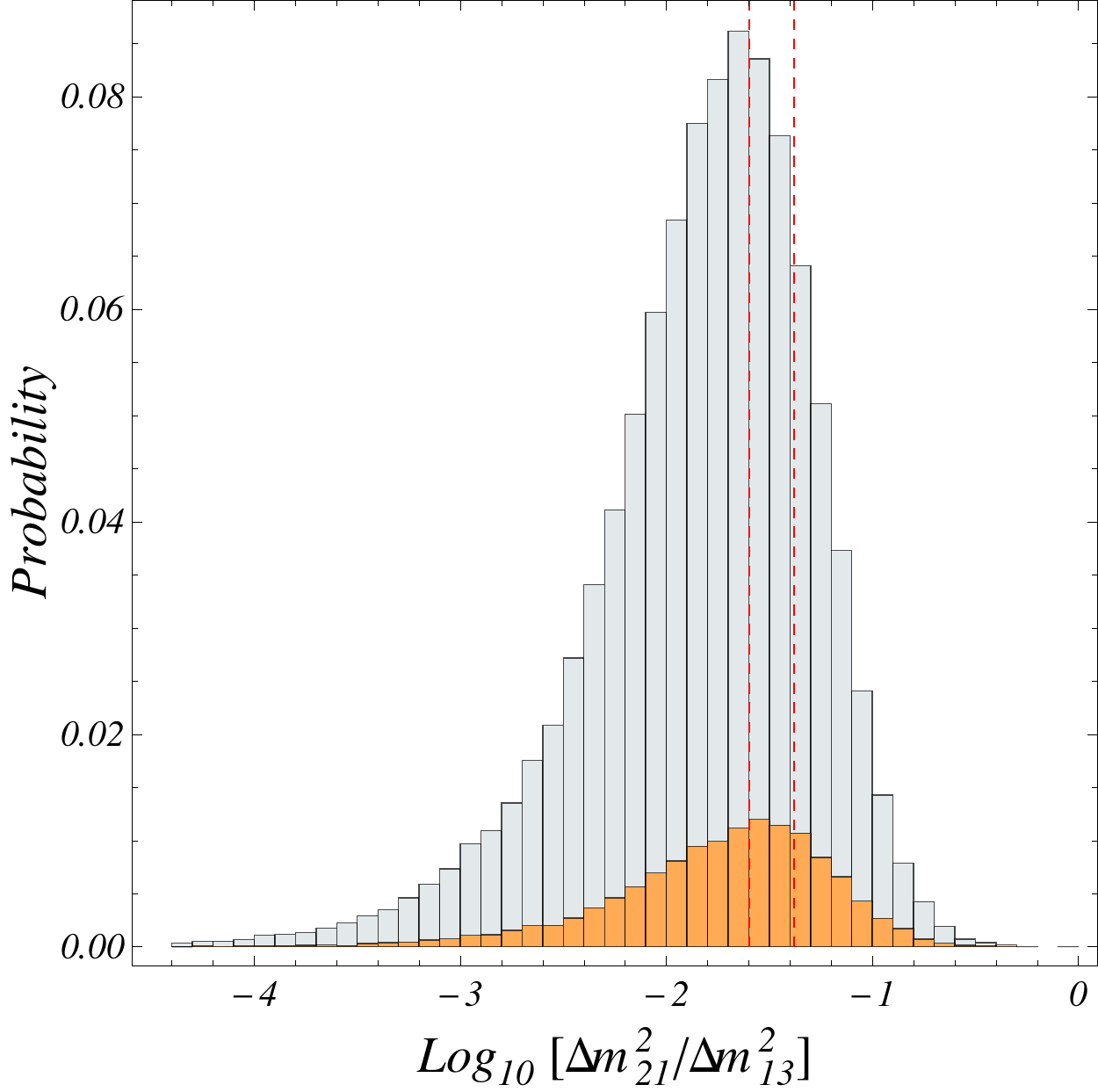}\\
\end{tabular}
\caption{\small \it 
{\rm \underline{Inverted hierarchy}} :  
Distributions of $\tan^2 \theta_{12}$, $\tan^2 \theta_{23}$, $\sin^2 \theta_{13}$ and
$\Delta m^2_{21}/\Delta m^2_{13}$ when the neutrino mass
matrix has the Cabibbo structure given by Eq.~(\ref{eq:ih}),
and with $U_l = U(13.0^\circ,0.2^\circ,2.4^\circ,0)$ ($U_l \simeq V_{CKM}$). 
In orange color : the distributions of the subsample ($13.5\%$ of total sample) for which 
$0.37 < \tan^2 \theta_{12} <0.61$ and $0.5 < \tan^2 \theta_{23} < 1.9$.
All other parameters are as in Fig.~1.}
\label{fig:ih}
\end{center}
\end{figure*}
Statistically, the structure Eq.~(\ref{eq:ih}) with an inverse hierarchy is significantly 
better to accommodate the experimental data (see Fig.~\ref{fig:ih}).
The probability that both angles $\theta_{12}$ and $\theta_{23}$ are large
and within the experimental range at $3\sigma$ C.L. is around $13.5\%$.
Interestingly, the Cabibbo angle $\theta_C = 13^\circ$ in $U_l$ ($\simeq V_{CKM}$) shifts the approximately 
maximal angle $\phi_{12}$ in $U_\nu$ to give the observed value 
$\theta_{12} \simeq \phi_{12} - \theta_C \simeq 32^\circ$.
Therefore the distribution shows a large probability around the observed value, 
and a very low probability of being maximal. 
In other words, this numerical study shows explicitly the validity of the so-called 
quark-lepton complementarity relation~(see \eg \cite{Minakata:2004xt}).
The biggest constraint is concerning the angle $\theta_{23}$, which is not predicted to be maximal
without an additional symmetry. 
On the positive side, the angle $\theta_{13}$ is automatically small for
the structure Eq.~(\ref{eq:ih}), with a probability $\sin^2 \theta_{13} < 0.05$ larger than $95\%$.
Also, the ratio $\Delta m^2_{21}/\Delta m^2_{13}$ is in a tighter range $10^{-3}-10^{-1}$ than in
the normal hierarchy case. Interestingly, when the two large angles are within the experimental range,
the third angle is small, but non zero, with a probability of $98.7\%$ that $0.01 < \sin^2 \theta_{13} < 0.04$.\\

\underline{Degenerate case} \\

With the structure Eq.~(\ref{eq:dh}) derived by supposing that the three light neutrinos are quasi degenerate 
in mass, it is very hard to accommodate the experimental data.
We obtain a probability around $0.025\%$ to have the two angles $\theta_{12}$ and $\theta_{23}$
large and within the experimental range at $3\sigma$ C.L.
Also, the observed hierarchy in the mass squared differences $\Delta m^2_{12}$ and $\Delta m^2_{13}$  is not
constrained by the mass matrix structure as $\Delta m^2_{12}/\Delta m^2_{13}$ has a large probability to be close to $1$.
We notice that a completely democratic neutrino mass matrix with no Cabibbo structure is statistically better
at explaining the data, as it gives a probability around $1.6\%$ to have the two large angles 
$\theta_{12}$ and $\theta_{23}$ within the experimental range at $3\sigma$ C.L.

\subsection{$U_l$ contains one large angle}
\label{sec:num2}

Here, as a maximal angle is automatically attributed to the charged lepton sector in the $2-3$ block, 
the statistical significance of having two large angles in $U_{MNS}$ is expectedly higher.
The purpose here is to compare different mass structures in the neutrino sector for a given
situation in the charged lepton sector.\\

\underline{Normal hierarchy}\\

\begin{figure*}[t]
\begin{center}
\begin{tabular}{ccc}
\includegraphics[width=0.45 \textwidth, clip]{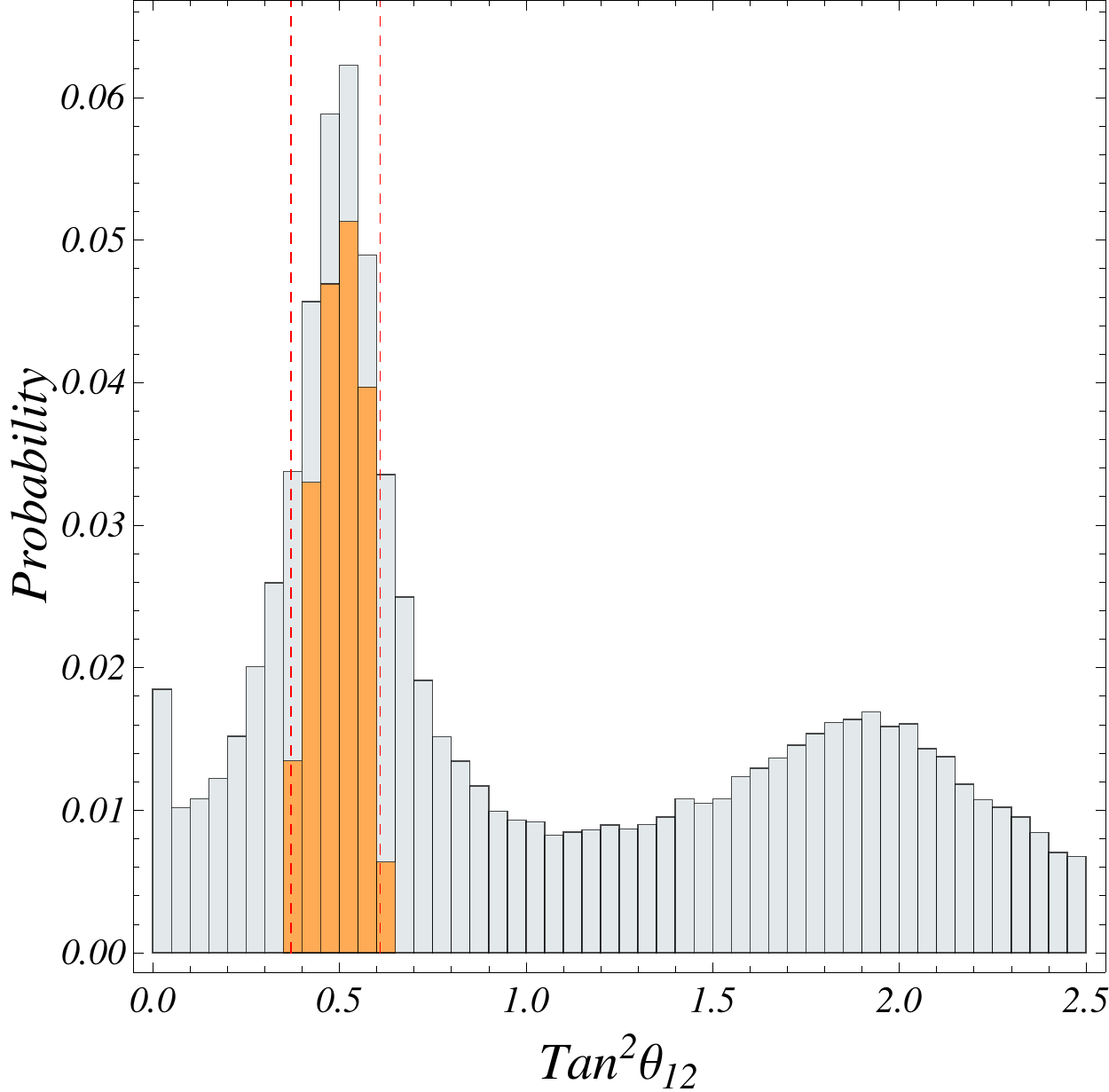}& ~ &
\includegraphics[width=0.45 \textwidth, clip]{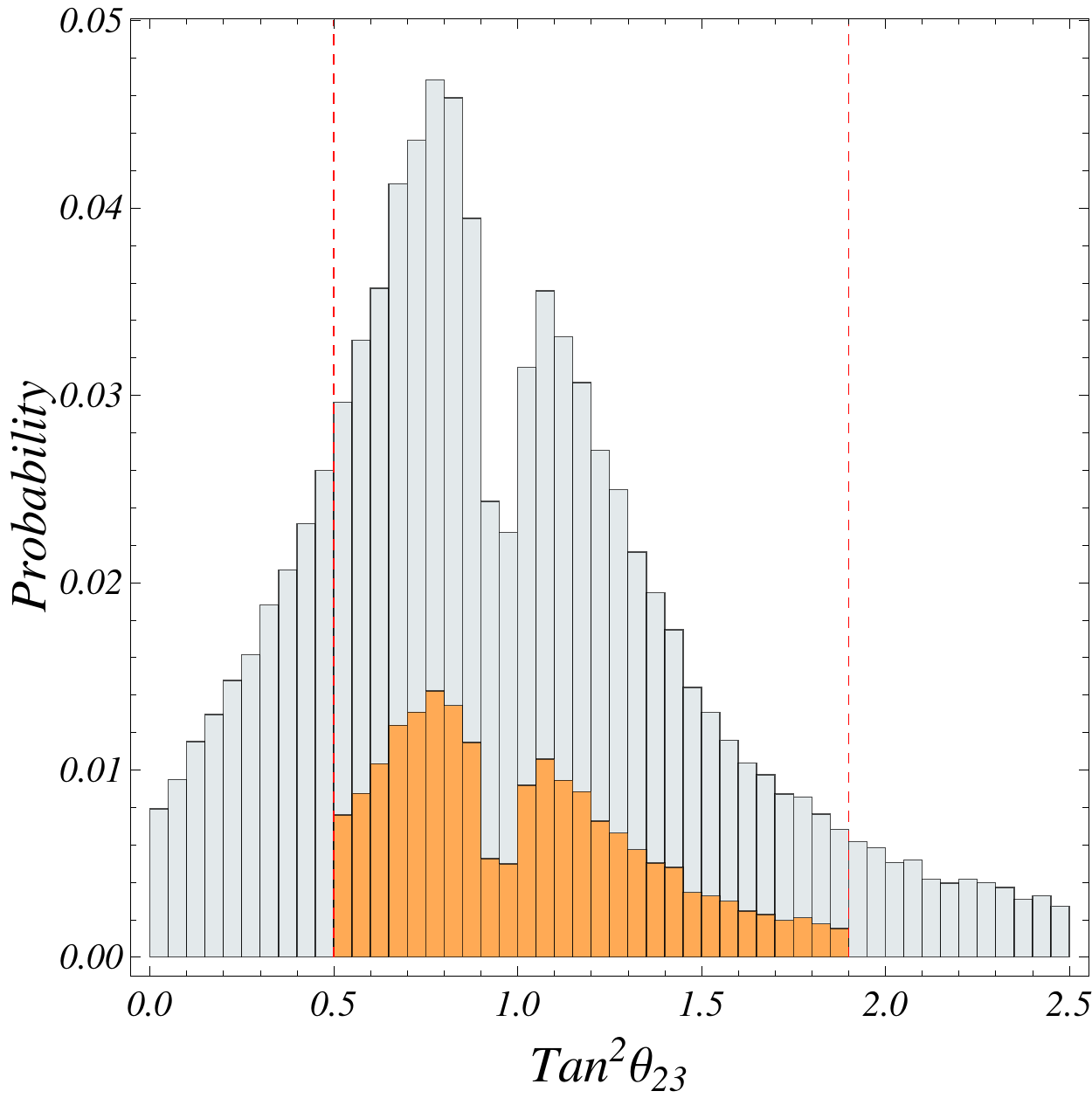}\\
\includegraphics[width=0.45 \textwidth, clip]{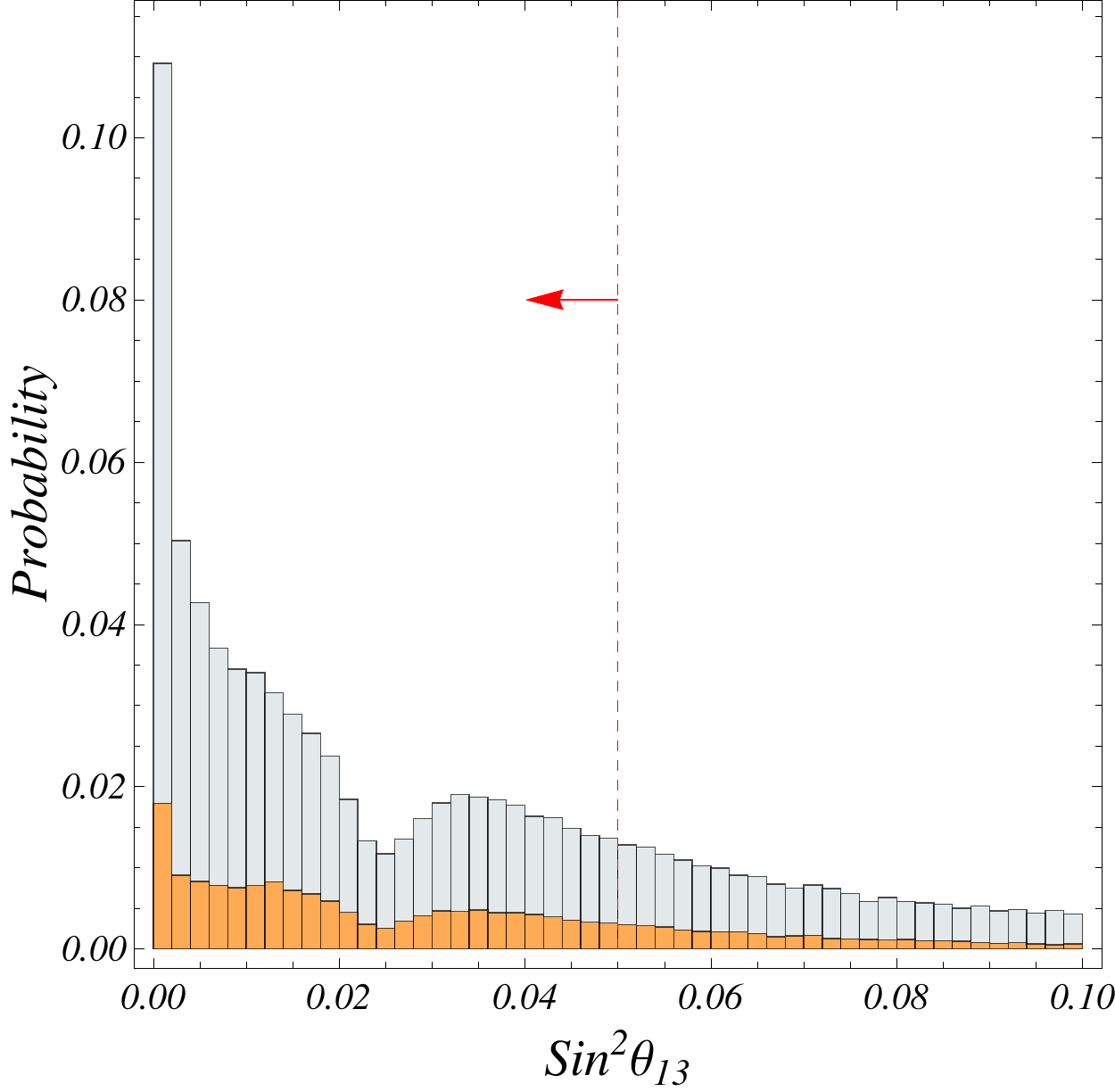}& ~ &
\includegraphics[width=0.45 \textwidth, clip]{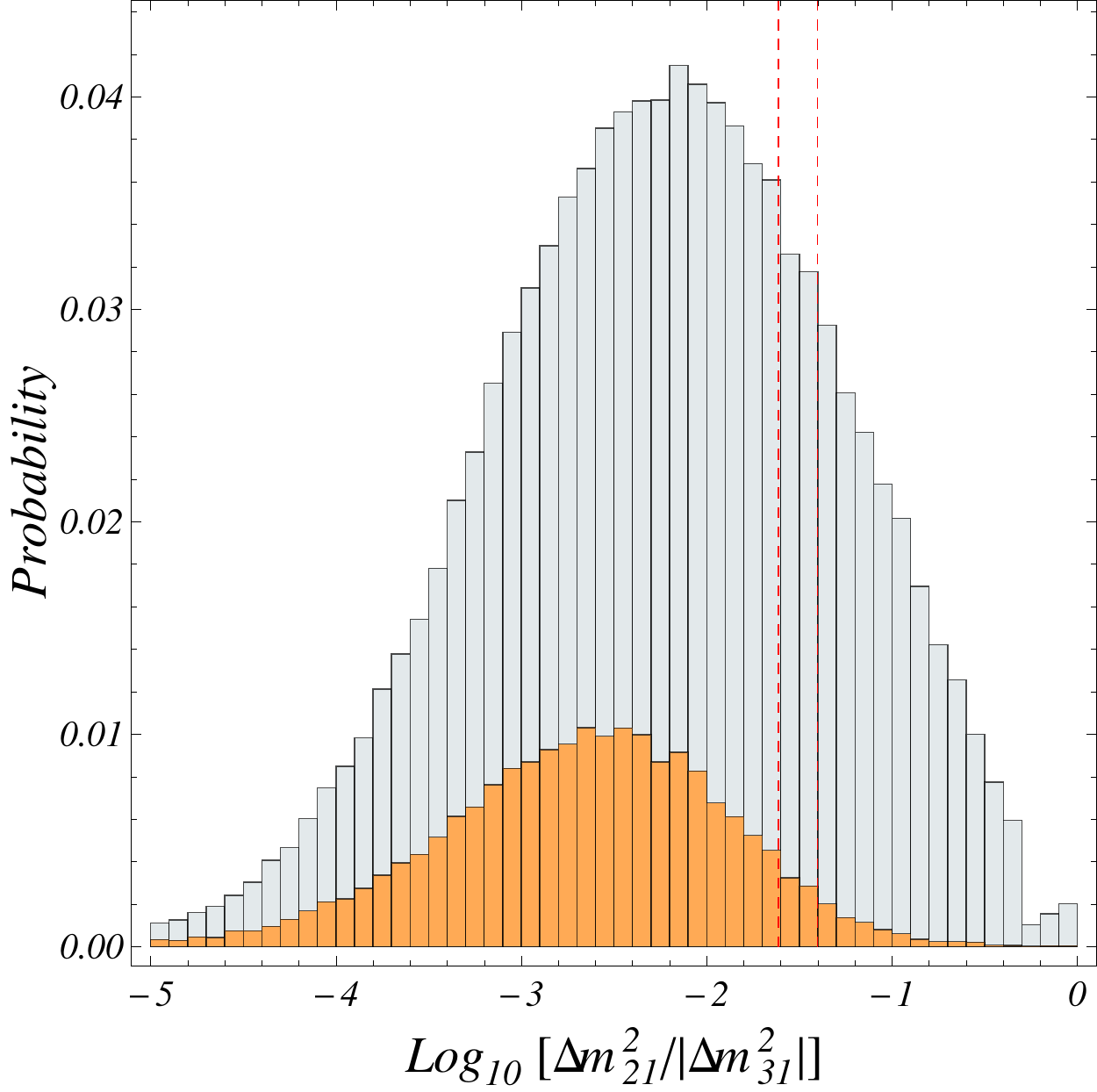}\\
\end{tabular}
\caption{\small \it 
{\rm \underline{Normal hierarchy}} :  
Distributions of $\tan^2 \theta_{12}$, $\tan^2 \theta_{23}$, $\sin^2 \theta_{13}$ and
$\Delta m^2_{21}/|\Delta m^2_{31}|$ when the neutrino mass matrix has the 
Cabibbo structure given by Eq.~(\ref{eq:nh3}),
and with $U_l = U(13.0^\circ,0.2^\circ,45.0^\circ,0)$ ($U_l$ contains a maximum angle in the $2-3$ block). 
In orange color : the distributions of the subsample ($19.1\%$ of total sample) for which 
$0.37 < \tan^2 \theta_{12} <0.61$ and $0.5 < \tan^2 \theta_{23} < 1.9$.
All other parameters are as in Fig.~1.}
\label{fig:nh2}
\end{center}
\end{figure*}
With the structures Eqs.~(\ref{eq:nh2}), the probability to have two large angles in $U_{MNS}$ 
in agreement with the experimental data is $6.8\%$. 
As with the structure Eq.~(\ref{eq:nh}), the problem is that the distribution of $\theta_{12}$ 
is extended and not centered around the observed value $\tan^2 \theta_{12} \simeq 0.5$.
In this respect, the structures Eqs.~(\ref{eq:nh3}) give a much better result,
with a probability to have two large angles around $19.1\%$ (see Fig.~\ref{fig:nh2}).
They give a hierarchical mass spectrum, but with the two light states forming a pseudo-Dirac pair.
As a result, they give a distribution of $\Delta m^2_{21}/|\Delta m^2_{31}|$ that
is more shifted towards smaller values.
Finally, we notice that for $1\%$ of the sample, the hierarchy is inverted rather than normal,
when the coefficient of the leading element in the matrix is accidentally small.\\

\underline{Inverted hierarchy} \\

\begin{figure*}[t]
\begin{center}
\begin{tabular}{ccc}
\includegraphics[width=0.45 \textwidth, clip]{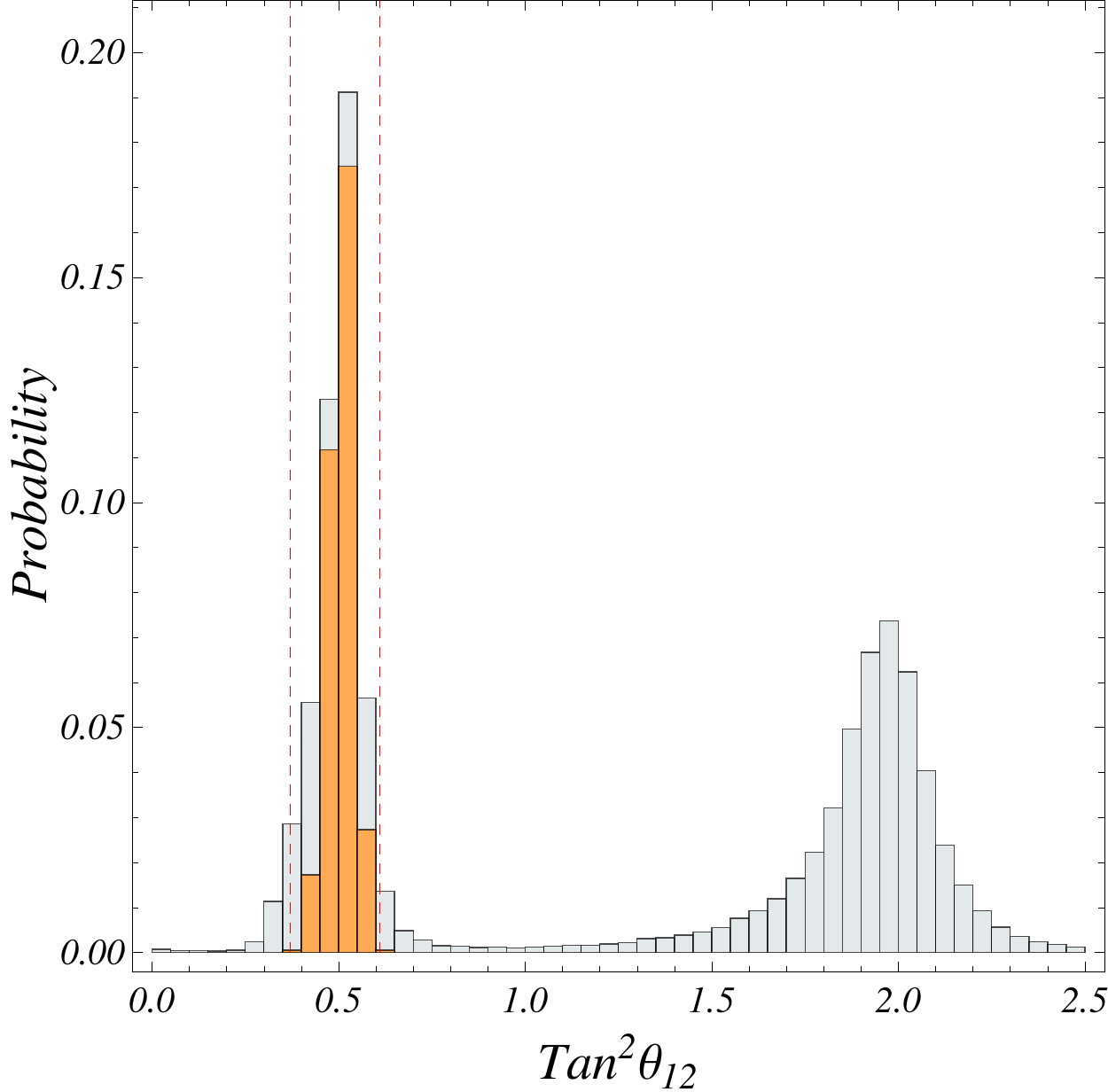}& ~ &
\includegraphics[width=0.45 \textwidth, clip]{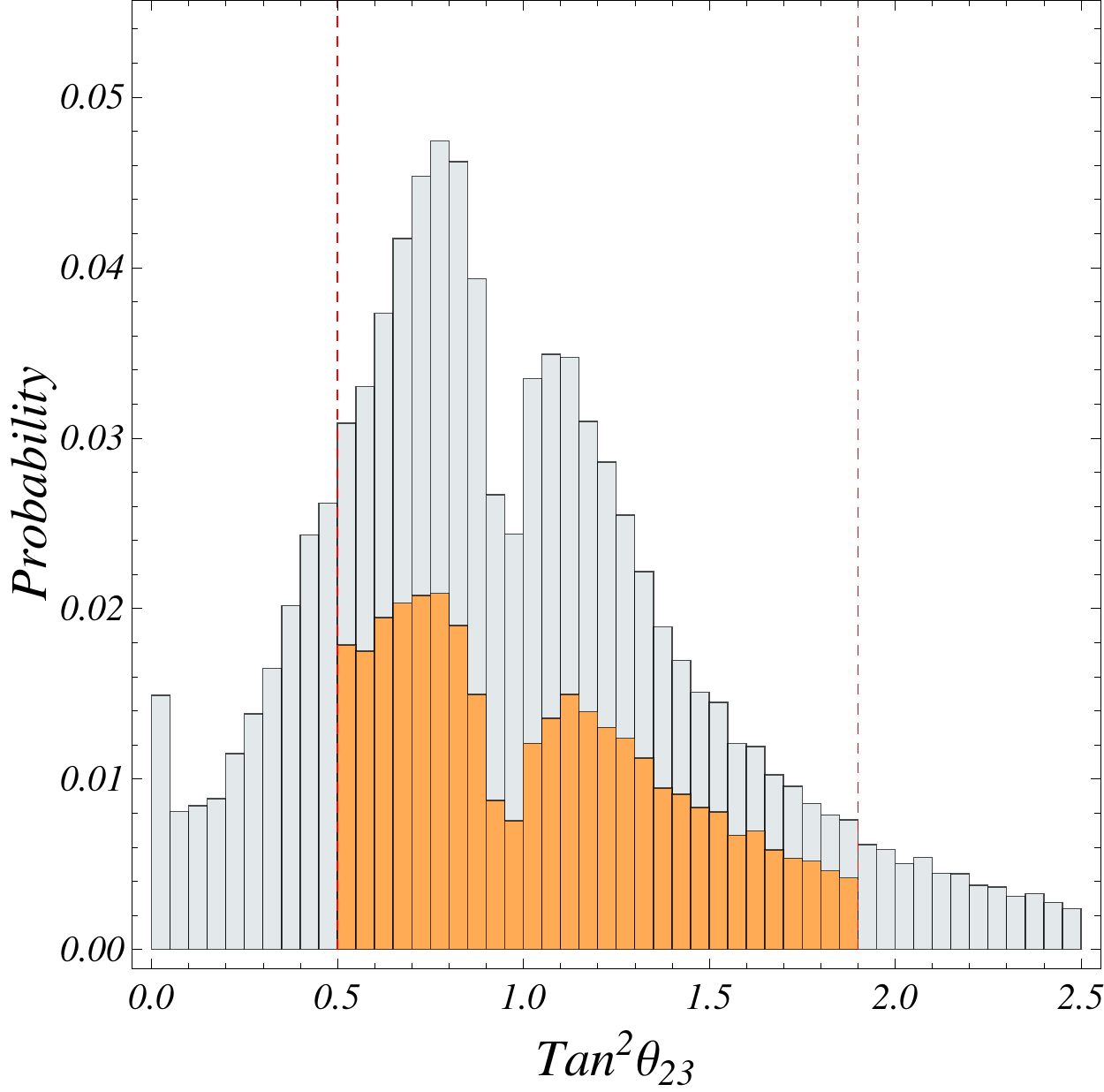}\\
\includegraphics[width=0.45 \textwidth, clip]{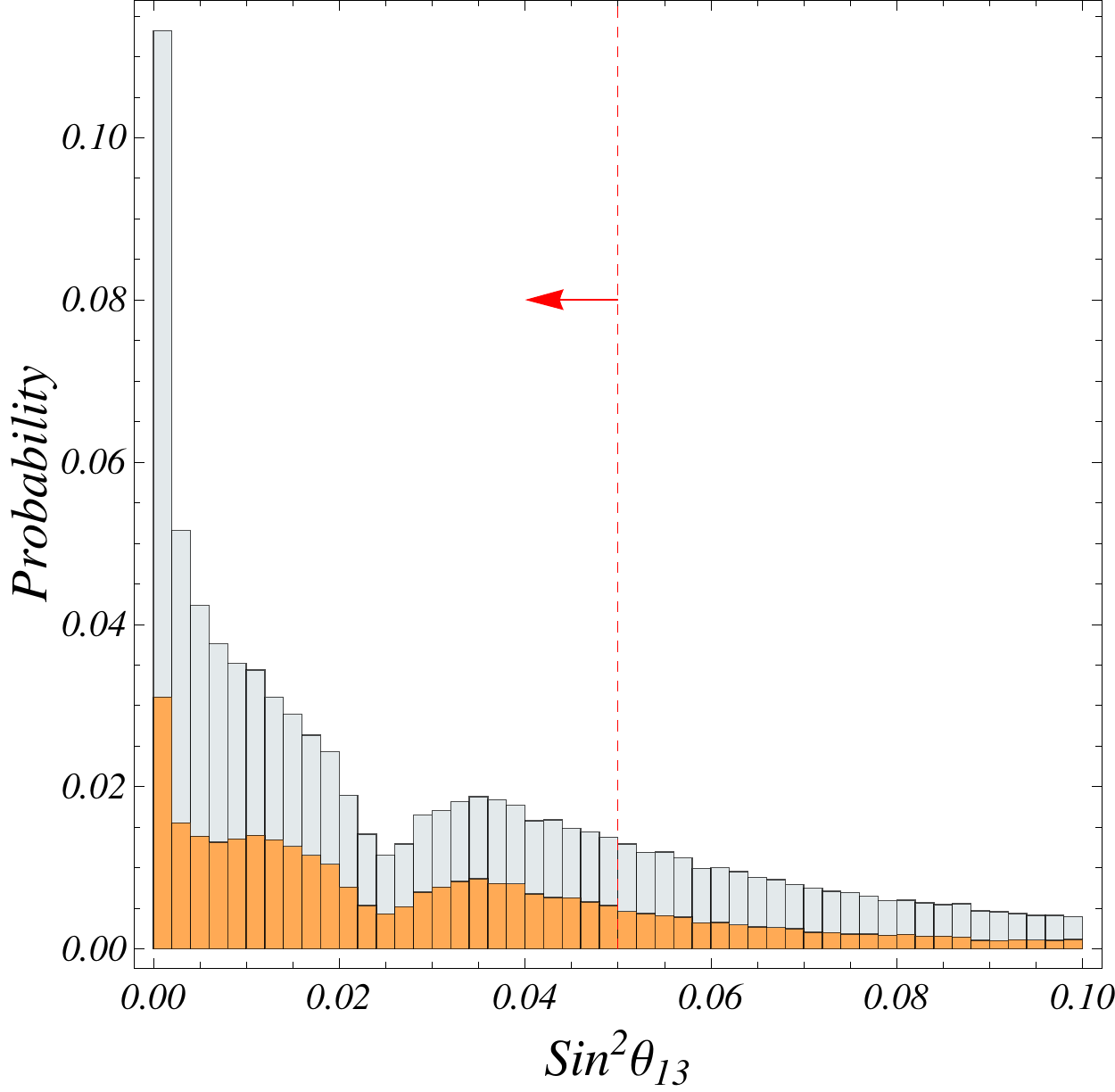}& ~ &
\includegraphics[width=0.45 \textwidth, clip]{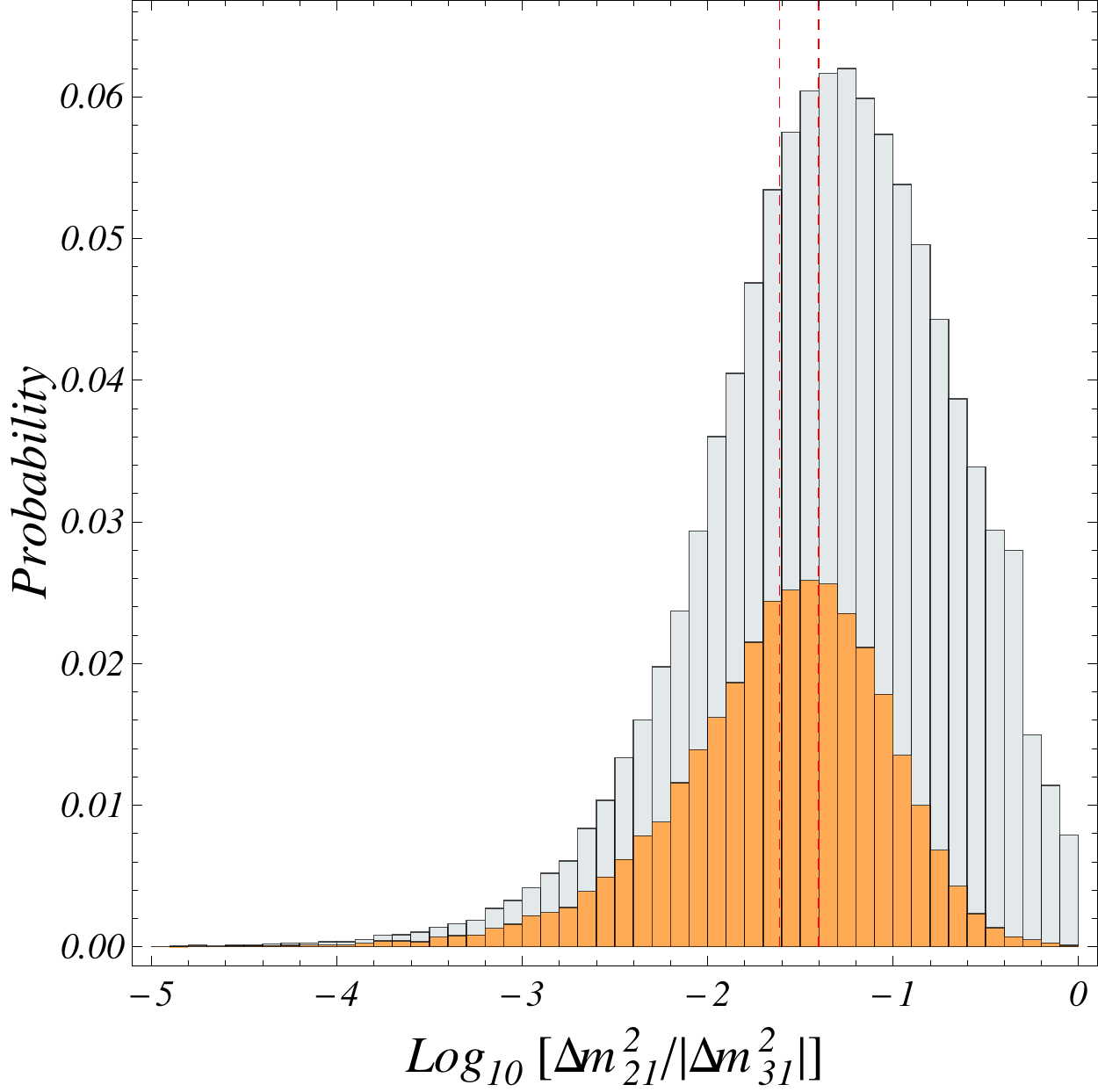}\\
\end{tabular}
\caption{\small \it 
{\rm \underline{Inverted hierarchy}} :  
Distributions of $\tan^2 \theta_{12}$, $\tan^2 \theta_{23}$, $\sin^2 \theta_{13}$ and
$\Delta m^2_{21}/|\Delta m^2_{13}|$ when the neutrino mass
matrix has the Cabibbo structure given by Eq.~(\ref{eq:ih2}),
and with $U_l = U(13.0^\circ,0.2^\circ,45.0^\circ,0)$ ($U_l$ contains a maximum angle in the $2-3$ block). 
In orange color : the distributions of the subsample ($33.4\%$ of total sample) for which 
$0.37 < \tan^2 \theta_{12} <0.61$ and $0.5 < \tan^2 \theta_{23} < 1.9$.
All other parameters are as in Fig.~1.}
\label{fig:ih2}
\end{center}
\end{figure*}
Next we consider the structures Eq.~(\ref{eq:ih2}), for which the pseudo-Dirac pair forms the heavier neutrino 
states, hence giving an inverted hierarchy (except for $1.6\%$ of the sample where the hierarchy is accidentally normal).
As a result, the angle $\theta_{12}$ is much less affected by subdominant Cabibbo effects,
leading to two very clean peaks around $\tan^2 \theta_{12} = 0.5$ and $\tan^2 \theta_{12} = 2.0$,
corresponding to $\theta_{12} = \pi/4 \pm \theta_C$.
The probability to have two large angles in the experimental range is substantially higher, at $33.4\%$.
Also, the distribution of $\Delta m^2_{21}/|\Delta m^2_{13}|$ is more peaked around the experimental
value. For $\sin^2 \theta_{13}$ however, there is no real prediction, the values cover
a large range from $0$ to beyond $0.1$. For the sub-sample with two large angles,
the probability that $\sin^2 \theta_{13}>0.05$ is $25.1\%$.\\

\underline{Degenerate case} \\

As discussed in Sec.~\ref{sec:23}, the structures Eq.~(\ref{eq:dh2}) do not in general give rise to a completely 
degenerate mass spectrum.
Instead, we have $m_{\nu_1} \simeq - m_{\nu_2} \sim m_{\nu_3}$, with an order of magnitude equality only 
for the last two neutrinos.
Numerically, we obtain a normal (inverted) spectrum $51\%$ ($49\%$) of the time.
The distributions for the angles and $\Delta m^2$ ratio are very similar to those of the 
inverted hierarchy case (Fig.~\ref{fig:ih2}), and therefore we do not display them.
The probability to have two large angles in the experimental range is $30.4\%$.

\section{Theoretical perspectives}
\label{sec:theory}

From the numerical study, we can already conclude that having dominant off-diagonal elements in the neutrino mass matrix
is the key to give rise systematically (\ie in a way that does not depend on a conspiracy of $\calo(1)$ numerical factors) 
to large angles in $U_{MNS}$. Also, the inverted hierarchy spectrum is significantly favored, as it leads to a more
accurate determination of the $\Delta m^2$ ratio, and a higher probability to have two large angles. 

The question here is how the various structures derived in Sec.~\ref{sec:data} can be constructed from a 
reasonable theory ? In particular, can these structures emerge in a unified flavor context 
where other fermions have hierarchical mass patterns ? The challenge is to understand how and why 
neutrinos seem to be in conflict with mass and mixing data from the other Standard Model fermions:
steep mass hierarchies for quarks and charged leptons, mild hierarchy of $\Delta m^2$ for neutrinos,
small mixings in the quark sector, two large mixings in neutrino oscillations. 

One very reasonable theory for neutrino masses is the well-known seesaw mechanism. 
It relates the tininess of neutrino Majorana masses to a very high energy scale where lepton number 
violating processes take place. Also, it offers potential clues to generate large mixing angles.
So we can re-ask the question in this context, whether the various structures derived in Sec.~\ref{sec:data} 
can be obtained with a seesaw mechanism. 
For instance, let us consider type I seesaw with three heavy Majorana neutrinos, so that the light neutrinos 
seesaw mass matrix is given by $M_\nu = M_D M_R^{-1} M_D^t$, where $M_D$ is the Dirac mass between the SM left-handed 
neutrinos and the heavy right-handed neutrinos, and $M_R$ is the Majorana mass of the right-handed neutrinos.

In the normal hierarchy case, a structure like Eq.~(\ref{eq:nh}) with a vanishing subdeterminant $D_1 \sim \lambda$
can easily be constructed in the case of a single right-handed neutrino dominance~\cite{King:2002nf}.
By this, we mean that the condition on $D_1$ which might appear as a fine-tuning requirement for the numerical prefactors, 
can actually be achieved naturally in the context of the seesaw mechanism.
Indeed, it suffices to take a diagonal Majorana mass matrix with very hierarchical masses $M_1 \ll M_2 \ll M_3$, 
so that the structure of the seesaw matrix is determined by $M_1$ alone, with the contributions coming 
from $M_2$ and $M_3$ being negligible.

This conclusion can even be extended in a rather straightforward although tedious way to the case 
where the Dirac mass matrix is hierarchical, \emph{i.e.} has hierarchical eigenvalues and small mixing angles.
The hierarchies in $M_D$ and $M_R$ can ``cancel" each other when they are somehow correlated~\cite{Datta:2003qg}, 
leading to a neutrino seesaw matrix with a large mixing angle and a vanishing 
subdeterminant $D_1$ in the $2-3$ block at leading order.
To see this, let's diagonalize $M_D$, $M_D = U_D \tilde{M_D} V_D^\dagger$, and suppose that
$\tilde{M_D} \sim m \, {\rm diag}(\lambda^{\alpha+\beta},\lambda^\alpha,1)$ with $\alpha, \beta > 0$,
$M_R^{-1} = M^{-1} \, {\rm diag}(1,\lambda^\gamma,\lambda^{\gamma+\delta})$ with $\gamma, \delta > 0$,
and that $U_D \sim V_D$ contain only small mixing angles
\be
U_D \sim V_D \sim \left(
\ba{ccc}
    1 & \lambda^{\alpha_1} & \lambda^{\alpha_3} \\
    \lambda^{\alpha_1} & 1 & \lambda^{\alpha_2} \\
    \lambda^{\alpha_3} & \lambda^{\alpha_2} & 1
\ea
\right) \quad .
\label{eq:vd}
\ee
For simplicity, we will suppose that $\alpha_3 = \alpha_1 + \alpha_2$.
Then, a large angle with a vanishing subdeterminant in the $2-3$ block of the seesaw matrix 
$M_\nu$ can be achieved in two cases. Either 
\be
\gamma < 2 \alpha_1 \quad , \quad 
\alpha = \alpha_2 \quad , \quad 
\delta > 2 \alpha_2 \quad , \quad 
\gamma < 2 \beta \quad , \quad
\ee
or
\be
\gamma > 2 \alpha_1 \quad , \quad 
\alpha = \alpha_2 \quad , \quad 
\gamma + \delta > 2 \alpha + 2 \alpha_1 \quad , \quad 
\beta > \alpha_1 \quad . \quad
\ee
For example, the values $\alpha = \beta = \alpha_1 = \alpha_2 = \gamma = 1$ and $\delta = 3$
lead to the structure of Eq.~(\ref{eq:nh}) with a subdeterminant $D_1 \sim \lambda$.

However, in this framework, it is not possible to get dominant off-diagonal elements in the neutrino seesaw matrix, 
without precise correlations between an angle $\zeta$ in the rotation matrix $V_D$ and the ratio of two of the Majorana masses,
say $M_a$ and $M_b$, like 
\be
\tan^2 \zeta =-\frac{M_a}{M_b}(1+\calo(\lambda^\alpha)) \quad ,
\label{eq:corr}
\ee
for some value $\alpha>0$ depending on the hierarchy of the Dirac masses. All possible cases have been examined 
in Ref.~\cite{Datta:2003qg}.
The condition Eq.~(\ref{eq:corr}) is a fine-tuning requirement except for the special case $M_a = -M_b$, 
as it corresponds to having a Dirac mass for the pair of right-handed neutrinos $N_{R,a}$ and $N_{R,b}$. 
So, as before, let us take $\tilde{M_D} \sim m \, {\rm diag}(\lambda^{\alpha+\beta},\lambda^\alpha,1)$ 
with $\alpha, \beta > 0$, $V_D$ given by Eq.~(\ref{eq:vd}), but let us consider a Majorana mass matrix with
a Dirac pattern in the $1-2$ block (the other cases involving the $1-3$ block are of course similar)
\be
M_R^{-1} = M^{-1} \left(
\ba{ccc}
    0 & 1 & 0 \\
    1 & 0 & 0 \\
    0 & 0 & \lambda^{\gamma}
\ea 
\right) \quad ,
\ee
with $\gamma > 0$. This can lead to a neutrino seesaw matrix with a pseudo-Dirac pattern in the $1-2$ block
if
\be
\beta < \alpha_1 \quad , \quad 
\alpha_2 \geq \alpha \quad , \quad 
\alpha_3, \, \alpha_2 + \gamma \geq \alpha + \beta \quad .
\ee
For example, the values $\alpha = \beta = \alpha_2 = 1$ and $\alpha_1 = \alpha_3 = \gamma = 2$
lead to the first structure of Eq.~(\ref{eq:nh3}).
The hierarchy type of the light neutrinos masses depends on the value of $\gamma$.
When $\gamma < 2 \alpha + \beta$, the hierarchy is normal, as in the previous example.
When $\gamma > 2 \alpha + \beta$, the hierarchy becomes inverted. We also need stricter inequalities
\be
\beta < \alpha_1 \quad , \quad 
\alpha_2 > \alpha \quad , \quad
\alpha_3 > \alpha + \beta \quad , \quad
\gamma > 2 \alpha + \beta \quad .
\ee
For example, the values $\alpha = \beta = 1$, $\gamma = 4$, $\alpha_2 = 2$ and $\alpha_1 = \alpha_3 = 3$
lead to the first structure of Eq.~(\ref{eq:ih2}).
When $\alpha_2 = \alpha$, it is possible to obtain an inverted hierarchy with two large 
angles. For example, the values $\alpha = \beta = 1$, $\gamma = 5$, $\alpha_1 = 3$, $\alpha_2 = 1$  and $\alpha_3 = 4$ lead to the structure Eq.~(\ref{eq:ih}).
Therefore, all the structures that we derived in Sec.~\ref{sec:data} can be obtained within the framework of the seesaw mechanism, and with the hypothesis that the neutrino Dirac mass matrix has hierarchical eigenvalues and small mixing angles as for the other Standard Model fermions.
However, a structure with dominant off-diagonal elements can be obtained without fine-tuning only if the Majorana mass matrix has itself a Dirac-type structure. One might of course question the origin of such a strange pattern in the right-handed neutrino sector.

One possible answer to this question is through extra-dimensions.
For instance, in six dimensions, a Majorana mass term always connects components with different 6D chiralities.
We can label components of a six dimensional Dirac field $\Psi$ according to their sign under
both $\Gamma_7$ and $\tilde{\Gamma}_5 = i \Gamma_0 \dots \Gamma_3$, with the
left- and right-handed chirality in 4D given by the projectors $P_{R,L} = (1 \mp \tilde{\Gamma}_5)/2$,
\be
\Psi = \left(
\ba{c}
\psi_{+R} \\ \psi_{+L} \\ \psi_{-L} \\ \psi_{-R}
\ea
\right) \quad .
\ee
So a 6D Dirac spinor is equivalent to two right-handed and two left-handed 4D Weyl spinors.
The Majorana mass term $\bar{\Psi}^c \Psi + {\rm h.c.}$ where $\Psi^c = C \Gamma^0 \Psi^*$,
and the charge conjugation operator is given by $C = \Gamma_0 \Gamma_2 \Gamma_4$ (up to a phase), becomes
\be
2\psi_{-R} \psi_{+R} - 2\psi_{-L} \psi_{+L} \quad + {\rm h.c.} 
\ee
where we use the contracted product notation
$\psi_R \psi_R \equiv \psi_R^t (i\sigma_2) \psi_R = \epsilon^{ij} \psi_{R i} \psi_{R j}$
for right-handed spinors and
$\psi_L \psi_L \equiv \psi_L^t (-i\sigma_2) \psi_L = -\epsilon^{ij} \psi_{L i} \psi_{L j}$
for left-handed spinors, where $\epsilon^{ij}$ is the totally antisymmetric tensor of rank 2.
We recall that $\psi \phi = \phi \psi$ as fermions anticommute.
Therefore, in a 6D theory, if a bulk gauge singlet $\Psi$ is embedded with a Majorana mass term, 
it will be give rise in the 4D point of view to a tower of massive states whose Majorana mass matrix is
totally off-diagonal. In Ref.~\cite{Frere:2010ah}, we demonstrate that a 6D theory with a vortex background gives a compelling 
solution to the flavor puzzle. 
The interaction of a single fermion in six dimensions with the vortex leads to three chiral zero-modes
in four dimensions, which might explain the replication of families in the Standard Model.
The different profiles of the three zero-modes can account for the quarks and charged lepton hierarchical masses.
In the neutrino sector, the off-diagonal structure of the Majorana mass explains the appearance of large mixing angles
through the seesaw mechanism. The structure of the fermion chiral zero-modes leads automatically to
a neutrino seesaw matrix with dominant off-diagonal elements, and hence a pseudo-Dirac inverted hierarchy mass spectrum.

\section{Summary \& Conclusions}
\label{sec:end}

In this paper, we have adopted a bottom-up approach to reconstruct the neutrino
 mass matrix from the neutrino oscillation data. We suppose that the neutrino masses are of Majorana type, and we take into account a possible rotation from the charged lepton sector. The possible structures are reconstructed using $\lambda \sim \lambda_C = \sin \theta_C$ as an expansion parameter.
 
We then perform a statistical analysis of the different structures using $\calo(1)$ numerical prefactors. It turns out that the presence of dominant off-diagonal elements is the key ingredient to explain the appearance of large angles in $U_{MNS}$. Also, structures with an inverted hierarchy are significantly favored over a normal or a degenerate hierarchy.
In other words, if we suppose that the neutrino (and charged lepton) mass matrices are not constrained by a special flavor symmetry, \eg a dicrete flavor symmetry based on a group $A_4$, $S_4$, $T'$ or $Z_n$, then structures with an inverted hierarchy like Eq.~(\ref{eq:ih}), Eqs.~(\ref{eq:ih2}) are largely favored.
The structure Eq.~(\ref{eq:ih}) is particularly interesting as it naturally leads to two large mixing angles in $U_\nu$, with an automatic suppression of the third angle $\theta_{13}$ below the experimental bound.

In Sec.~\ref{sec:num}, we tried to address two questions. The first one is whether the neutrino mass matrix structures derived in Sec.~\ref{sec:data} can or cannot be obtained in a reasonnable neutrino theory. The second one is whether the neutrino data, especially the presence of two large mixing angles, can or cannot be reconciled with the hierarchical masses and small mixing angles found in the other sectors of the Standard Model. We show that within the framework of the seesaw mechanism, both questions can be positively answered, and we explicitly reconstruct the structures considered in Sec.~\ref{sec:data}. The bottomline is that the seesaw mechanism can be responsible for the appearance of large angles in the neutrino sector, but only at the puzzling condition that the Majorana mass matrix for the right-handed heavy neutrinos itself bears a Dirac-type pattern for two states. We show that this last condition becomes natural in six-dimensional theories, because a Majorana mass term in 6D always connects components with different 6D chiralities, so that in the 4D point of view, the Majorana mass matrix of the tower of massive states is totally off-diagonal.
 
\acknowledgments

I would like to thank Jean-Marie Fr\`ere for interesting discussions.
This work is funded in part by IISN and by Belgian Science Policy (IAP VI/11). 

\bibliographystyle{JHEP}
\bibliography{naturalnu}

\providecommand{\href}[2]{#2}\begingroup\raggedright\begin{thebibliography}{10}

\bibitem{Amsler:2008zzb}
{\bf Particle Data Group} Collaboration, C.~Amsler {\em et.~al.}, {\it {Review
  of particle physics}},  {\em Phys. Lett.} {\bf B667} (2008) 1.

\bibitem{GonzalezGarcia:2010er}
M.~C. Gonzalez-Garcia, M.~Maltoni, and J.~Salvado, {\it {Updated global fit to
  three neutrino mixing: status of the hints of theta13 $>$ 0}},  {\em JHEP}
  {\bf 04} (2010) 056, [\href{http://xxx.lanl.gov/abs/1001.4524}{{\tt
  arXiv:1001.4524}}].

\bibitem{Apollonio:2002gd}
{\bf CHOOZ} Collaboration, M.~Apollonio {\em et.~al.}, {\it {Search for
  neutrino oscillations on a long base-line at the CHOOZ nuclear power
  station}},  {\em Eur. Phys. J.} {\bf C27} (2003) 331--374,
  [\href{http://xxx.lanl.gov/abs/hep-ex/0301017}{{\tt hep-ex/0301017}}].

\bibitem{Komatsu:2010fb}
E.~Komatsu {\em et.~al.}, {\it {Seven-Year Wilkinson Microwave Anisotropy Probe
  (WMAP) Observations: Cosmological Interpretation}},
  \href{http://xxx.lanl.gov/abs/1001.4538}{{\tt arXiv:1001.4538}}.

\bibitem{Kraus:2004zw}
C.~Kraus {\em et.~al.}, {\it {Final Results from phase II of the Mainz Neutrino
  Mass Search in Tritium $\beta$ Decay}},  {\em Eur. Phys. J.} {\bf C40} (2005)
  447--468, [\href{http://xxx.lanl.gov/abs/hep-ex/0412056}{{\tt
  hep-ex/0412056}}].

\bibitem{Thomas:2009ae}
S.~A. Thomas, F.~B. Abdalla, and O.~Lahav, {\it {Upper Bound of 0.28 eV on the
  Neutrino Masses from the Largest Photometric Redshift Survey}},  {\em Phys.
  Rev. Lett.} {\bf 105} (2010) 031301,
  [\href{http://xxx.lanl.gov/abs/0911.5291}{{\tt arXiv:0911.5291}}].

\bibitem{Gatto:1968ss}
R.~Gatto, G.~Sartori, and M.~Tonin, {\it {Weak Selfmasses, Cabibbo Angle, and
  Broken SU(2) x SU(2)}},  {\em Phys. Lett.} {\bf B28} (1968) 128--130.

\bibitem{Cabibbo:1968vn}
N.~Cabibbo and L.~Maiani, {\it {Dynamical interrelation of weak,
  electromagnetic and strong interactions and the value of theta}},  {\em Phys.
  Lett.} {\bf B28} (1968) 131--135.

\bibitem{Froggatt:1978nt}
C.~D. Froggatt and H.~B. Nielsen, {\it {Hierarchy of Quark Masses, Cabibbo
  Angles and CP Violation}},  {\em Nucl. Phys.} {\bf B147} (1979) 277.

\bibitem{Ling:2002nj}
F.-S. Ling and P.~Ramond, {\it {Family hierarchy and large neutrino mixings}},
  {\em Phys. Lett.} {\bf B543} (2002) 29--37,
  [\href{http://xxx.lanl.gov/abs/hep-ph/0206004}{{\tt hep-ph/0206004}}].

\bibitem{Ramond:1979py}
P.~Ramond, {\it {The Family Group in Grand Unified Theories}},
  \href{http://xxx.lanl.gov/abs/hep-ph/9809459}{{\tt hep-ph/9809459}}.

\bibitem{Yanagida:1979as}
T.~Yanagida, {\it {Horizontal gauge symmetry and masses of neutrinos}}, . In
  Proceedings of the Workshop on the Baryon Number of the Universe and Unified
  Theories, Tsukuba, Japan, 13-14 Feb 1979.

\bibitem{King:2002nf}
S.~F. King, {\it {Constructing the large mixing angle MNS matrix in see-saw
  models with right-handed neutrino dominance}},  {\em JHEP} {\bf 09} (2002)
  011, [\href{http://xxx.lanl.gov/abs/hep-ph/0204360}{{\tt hep-ph/0204360}}].

\bibitem{Minakata:2004xt}
H.~Minakata and A.~Y. Smirnov, {\it {Neutrino Mixing and Quark-Lepton
  Complementarity}},  {\em Phys. Rev.} {\bf D70} (2004) 073009,
  [\href{http://xxx.lanl.gov/abs/hep-ph/0405088}{{\tt hep-ph/0405088}}].

\bibitem{Datta:2003qg}
A.~Datta, F.-S. Ling, and P.~Ramond, {\it {Correlated hierarchy, Dirac masses
  and large mixing angles}},  {\em Nucl. Phys.} {\bf B671} (2003) 383--400,
  [\href{http://xxx.lanl.gov/abs/hep-ph/0306002}{{\tt hep-ph/0306002}}].

\bibitem{Frere:2010ah}
J.-M. Fr\`ere, M.~Libanov, and F.-S. Ling, {\it {See-saw neutrino masses and
  large mixing angles in the vortex background on a sphere}},
  \href{http://xxx.lanl.gov/abs/1006.5196}{{\tt arXiv:1006.5196}}.

\end{thebibliography}\endgroup

\end{document}